\documentclass[a4paper,11pt]{article}
\usepackage{graphicx}
\usepackage{amscd}
\usepackage[all]{xy}
\usepackage{amsmath}
\usepackage{amssymb,amsfonts}
\usepackage{indentfirst,latexsym,bm,cite}

\newtheorem{thm}{Theorem}[section]
\newtheorem{lem}[thm]{Lemma}

\begin{document}
\title{Numerical calculation of N-periodic wave solutions to coupled KdV-type equations}

\author{Yingnan Zhang\textsuperscript{\,1}, Xingbiao Hu\textsuperscript{\,2,3} and Jianqing Sun\textsuperscript{\,4}\thanks{sunjq@lsec.cc.ac.cn}}

\date{}%
\maketitle
\begin{center}
\textsuperscript{\,1}Jiangsu Key Laboratory for NSLSCS, School of Mathematical Sciences, Nanjing Normal University, Nanjing, China\\
\textsuperscript{\,2}LSEC, Institute of Computational Mathematics
and Scientific Engineering Computing, Academy of Mathematics and Systems Sciences, Chinese Academy of
Sciences, Beijing, China\\
\textsuperscript{\,3}School of Mathematical Sciences, University of Chinese Academy of Sciences, Beijing, China\\
\textsuperscript{\,4}School of Mathematical Sciences, Ocean University of China, Qingdao, China\\
\end{center}

\begin{abstract}
In this paper, we study periodic wave solutions of coupled KdV-type equations.  We present a numerical process to calculate the $N$-periodic waves based on the direct method of calculating periodic wave solutions proposed by Akira Nakamura.  Particularly, in the case of $N=1,2,3$, we give some detailed examples to show the N-periodic wave solutions to a coupled Ramani equation.
\end{abstract}

\begin{quote}
\end{quote}

\section{Introduction}
\label{section1}
In this paper, we focus on  numerical calculation of  $N$-periodic wave solutions to coupled KdV-type soliton equations.
The periodic solution mentioned here represents periodic analogue of soliton solution, and in general case $N$-periodic wave solution is a periodic generalization of $N$-soliton solution or multiple collision of $N$ solitons\cite{nakamura1}.

Much work has already been done on periodic waves.  The pioneering work was made by Novikov and Dubrovin\cite{novikov,Dubrovin1,Dubrovin2}, Lax\cite{lax1975}, Its and Matveev\cite{Matveev}, McKean and Moerbeke\cite{Moerbeke} in 1970s.
After that, some classical methods such as the inverse scattering method\cite{ablowitz1981,forest1982,date}, the algebro-geometric approach\cite{alg1,alg2,alg3,alg4,alg5,alg6,alg7} and the direct method\cite{nakamura1,nakamura2,hirota2,fan1,fan2,fan3}, are applied to solve  periodic waves.  However, comparing with soliton waves, the periodic waves are more complicated and it is difficult to give some detailed explicit expressions. Therefore, many researchers turn to numerical calculations.  Recent work includes the numerical approach via Riemann-Hilbert problem\cite{rh1,rh2} and spectral method\cite{Klein2004,Klein2006,Klein2012}. Here we will calculate the periodic waves numerically based on the direct method\cite{hirota1,nakamura1,nakamura2,hirota2}.

In Refs.~\cite{nakamura1}-\cite{nakamura2}, Nakamura first proposed the conditions for having $N$-periodic waves  to nonlinear evolution equation which can be reduced to some certain type of bilinear equations, such as KdV, mKdV, NLS and some other equations. And then in Ref.~\cite{hirota2}, Hirota suggested researchers investigate whether the soliton equations written in bilinear form exhibit $3$-periodic wave solutions or not by this condition.

In this paper, we will apply the conditions to coupled KdV-type equations and give a numerical procedure to calculate their $N$-periodic wave solutions. Here ``coupled KdV-type" means that, with some suitable variable transformations and auxiliary variables,  equation can be transformed into the following bilinear form
\begin{eqnarray}
&&F_1(D_t,D_z,D_x,\cdots,c_1)f\cdot f=0,\label{ckdv1}\\
&&F_2(D_t,D_z,D_x,\cdots,c_2)f\cdot f=0,\label{ckdv2}
\end{eqnarray}
where $F_1,F_2$ are even functions of $D_t,D_z,D_x,\cdots,$  $c_1, c_2$ are integral constants, and the $D$ operator\cite{hirota1} is defined by
\begin{eqnarray}
& & D_t^mD_x^n a(t,x)\cdot b(t,x)\nonumber\\
&=&\frac{\partial^m}{\partial s^m}\frac{\partial^n}{\partial y^n}a(t+s,x+y)b(t-s,x-y)|_{s=0,y=0},\nonumber\\
&& m,n=0,1,2,\cdots.
\end{eqnarray}

Many  soliton equations can be viewed as coupled KdV-type equations. For example,  the coupled Ramani equation\cite{cramani1}
\begin{eqnarray}
&&u_{xxxxxx}+15u_{xx}u_{xxx}+15u_xu_{xxxx}+45u_x^2u_{xx}\nonumber\\
&&\quad-5(u_{xxxt}+3u_{xx}u_t+3u_xu_{xt})-5u_{tt}+18v_x=0,\label{cramani1}\\
&&v_t-v_{xxx}-3v_xu_x-3vu_{xx}=0, \label{cramani2}
\end{eqnarray}
can be transformed into the bilinear form
\begin{eqnarray}
&&(D_x^6-5D_x^3D_t-5D_t^2+9D_xD_z+c_1)f\cdot f=0,\label{cramani3}\\
&&(D_zD_t-D_zD_x^3-6v_0 D_x^2+c_2)f\cdot f=0,\label{cramani4}
\end{eqnarray}
by the dependent variable transformation
\begin{equation}
u=u_0+(\mathrm{ln} f)_{xx},\quad v=v_0+(\mathrm{ln}f)_{xz}, \label{vtran1}
\end{equation}
where $z$ is an auxiliary variable and $c_1,c_2$ are integral constants. This type of equations also include the Hirota-Satsuma coupled KdV equation\cite{hs},  the Camassa-Holm equation\cite{ch1,ch2},  the semi-discrete KdV equation\cite{hirota3} and some other discrete soliton equations\cite{zhang1,zhang2}.

In the case of single KdV-type bilinear equations, as shown by Nakamura\cite{nakamura1} and Hirota\cite{hirota2}, there are always exactly $1$- and $2$- periodic wave solutions and if $N\geq 3$, we need solve an over-determined nonlinear algebraic system to obtain a $N$-periodic wave.
However, the situation of  coupled KdV-type equations is different. In this case,  there are only exactly $1$-periodic wave solutions and if $N\geq 2$, it is necessary for us to deal with an over-determined algebraic system to solve $N$-periodic wave solutions( see Sect.~\ref{section2} for details).

The paper is organized as follows. In Sect.~\ref{section2}, we will review the conditions given by Nakamura and apply them to  the coupled KdV-type bilinear Eqs.~\eqref{ckdv1}-\eqref{ckdv2}.  In Sect.~\ref{section3},  we will propose a numerical procedure by using Gauss-Newton method based on the condition.  Section \ref{section4} devotes to some numerical results of the coupled Ramani Eqs.~ \eqref{cramani3}-\eqref{cramani4}. Some conclusions and discussions will be given in Sect.~\ref{section5}.

\section{Condition for $N$-periodic wave solutions}
\label{section2}
Firstly, we review the Riemann's $\theta$-function defined by
\begin{eqnarray}
&&\theta({\bm{\eta}};{\bm{s}}|{\bm{\tau}})=\sum_{m_1}\sum_{m_2}\cdots\sum_{m_{N}=-\infty}^{\infty}\exp[i\sum_{j=1}^{N}(m_j+s_j)\eta_j\nonumber\\
&&\qquad\qquad\qquad-\frac{1}{2}\sum_{j,k=1}^N(m_j+s_j)\tau_{j,k}(m_k+s_k)],\label{thetaf}
\end{eqnarray}
where $\eta_j$, $s_j$ and $\tau_{j,k}$ are the elements of the vector $\bm{\eta}$, $\bm{s}$ and the symmetric matrix $\bm{\tau}$ respectively, and $\eta_j$ is defined by
\begin{equation}
\eta_j=\omega_jt+k_jx+\cdots+\eta_j^0,\quad j=1,2,\cdots N.
\end{equation}
Here $k_j$, $\omega_j$, $\eta_j^0$, the diagonal elements $\tau_{jj}$ and off-diagonal elements $\tau_{jk}, j\neq k$ are parameters corresponding to the wave numbers, the frequencies, the phase positions, the amplitudes and the interactions respectively.

\subsection{Single KdV-type bilinear equations}
 For a single KdV-type bilinear equation
 \begin{equation}
 F(D_t,D_x,\cdots,\lambda)f\cdot f=0, \label{kdv1}
 \end{equation}
the condition for having $N$-periodic wave solutions was first proposed by Nakamura\cite{nakamura1}.
\begin{lem}
For the Riemann's $\theta$-function defined by \eqref{thetaf}, $\theta(\bm{\eta};\bm{0}|\bm{\tau})$ is a $N$-periodic wave solution of the single bilinear equation \eqref{kdv1} if
\begin{eqnarray}
&&\qquad\sum_{m_1}\sum_{m_2}\cdots\sum_{m_{N}=-\infty}^{\infty}\nonumber\\
&&F[2i\sum_{j=1}^{N}(m_j-\mu_j/2)\omega_j,2i\sum_{j=1}^{N}(m_j-\mu_j/2)k_j,\cdots,\lambda]\nonumber\\
&&\qquad\times\exp[-\sum_{j,k=1}^N(m_j-\mu_j/2)\tau_{j,k}(m_k-\mu_k/2)]=0,\label{con1}
\end{eqnarray}
for all possible combinations $\mu_1=0,1$, $\mu_2=0,1$, $\cdots$, $\mu_N=0,1$.
\end{lem}
The proof of this lemma can be found in Refs.~\cite{nakamura1,hirota2,osborne} .

Note that there are $2^N$ equations of type \eqref{con1}, and the total number of parameters $\omega_j,k_j(j=1,2,\cdots,N)$, $\lambda$, and $\tau_{j,k}(1\leq j,k\leq N)$ is $2N+1+N(N+1)/2$.  Generally, the diagonal elements $\tau_{jj}$ which influence the amplitudes, and the wave numbers $k_j$(or frequencies $\omega_j$) are taken to be given parameters. Thus we have $2^N$ equations with $1+N(N+1)/2$ unknowns.
In the case of $N=1,2$, we have the equal number of equations and unknown parameters while in the case of $N\geq3$, the number of equations is larger than the number of unknown parameters, which means that this is an over-determined system.

\subsection{Coupled KdV-type bilinear equations}
We apply Lemma~2.1 to coupled KdV-type bilinear Eqs.~\eqref{ckdv1}-\eqref{ckdv2}. Note that, there is an auxiliary variable $z$  in this system. Therefore, the $\eta_j$ in the Riemann's $\theta$-function \eqref{thetaf} is defined by
\begin{equation}
\eta_j=\omega_jt+l_jz+k_jx+\cdots+\eta_j^0,\quad j=1,2,\cdots N.\label{etaj}
\end{equation}
We have the following theorem.
\begin{thm}
For the Riemann's $\theta$-function defined by \eqref{thetaf}-\eqref{etaj}, $\theta(\bm{\eta};\bm{0}|\bm{\tau})$ is a $N$-periodic wave solution of the coupled bilinear equation \eqref{ckdv1}-\eqref{ckdv2} if
\begin{eqnarray}
&&\sum_{m_1}\sum_{m_2}\cdots\sum_{m_{N}=-\infty}^{\infty}F_1[2i\sum_{j=1}^{N}(m_j-\mu_j/2)\omega_j,\nonumber\\
&&2i\sum_{j=1}^{N}(m_j-\mu_j/2)l_j,2i\sum_{j=1}^{N}(m_j-\mu_j/2)k_j,\cdots,c_1]\nonumber\\
&&\quad\times\exp[-\sum_{j,k=1}^N(m_j-\mu_j/2)\tau_{j,k}(m_k-\mu_k/2)]=0,\label{con2}\\
&&\sum_{m_1}\sum_{m_2}\cdots\sum_{m_{N}=-\infty}^{\infty}F_2[2i\sum_{j=1}^{N}(m_j-\mu_j/2)\omega_j,\nonumber\\
&&2i\sum_{j=1}^{N}(m_j-\mu_j/2)l_j,2i\sum_{j=1}^{N}(m_j-\mu_j/2)k_j,\cdots,c_2]\nonumber\\
&&\quad\times\exp[-\sum_{j,k=1}^N(m_j-\mu_j/2)\tau_{j,k}(m_k-\mu_k/2)]=0,\label{con3}
\end{eqnarray}
for all possible combinations $\mu_1=0,1$, $\mu_2=0,1$, $\cdots$, $\mu_N=0,1$.
\end{thm}
The proof of this theorem is similar to that of Lemma~2.1 which is to substitute $\theta(\bm{\eta};\bm{0}|\bm{\tau})$ into the coupled bilinear system \eqref{ckdv1}-\eqref{ckdv2} and simplify the formula with some bilinear identities and tedious calculations. We omit the details here.

Note that there are $2^{N+1}$ equations, and the total number of parameters  $\omega_j,l_j,k_j(j=1,2,\cdots,N)$, $c_1,c_2$, and $\tau_{j,k}(1\leq j,k\leq N)$ is $3N+2+N(N+1)/2$. With $k_j$ and $\tau_{jj}$ given, we obtain a nonlinear algebraic system of $2^{N+1}$ equations with $N+2+N(N+1)/2$ unknowns. Thus, for $1$-periodic waves, we need to solve $4$ parameters from $4$ equations while for $2$- and $3$-periodic waves, we have to solve $7$ and $11$ parameters from $8$ and $16$ equations respectively.

\section{Numerical scheme}
\label{section3}
In this section, we will introduce our numerical procedure to solve the unknown parameters from the nonlinear algebraic system\eqref{con2}-\eqref{con3}. The main idea is to formulate the problem as a nonlinear least square problem and then use the Gauss-Newton method\cite{bjorck} to solve it.

For simplicity, we rewrite Eqs.~\eqref{con2}-\eqref{con3} as
\begin{equation}
{\bm{H(\bm{x}\bm)}}=(H_1,H_2,\cdots,H_{2^{N+1}})^T=0,
\end{equation}
where $H_i=0(i=1,2,\cdots,2^{N+1})$ is one of the equations in system \eqref{con2}-\eqref{con3} and
${\bm{x}}$ is a vector whose elements are the unknown parameters $\omega_j$, $l_j$, $\tau_{ij}(i<j)$ and $c_1,c_2$.
The objective function of the nonlinear least square problem is
\begin{equation}\label{objectivefun}
S({\bm{x}})=\frac{1}{2}\sum_{n=1}^{2^{N+1}}H_n^2({\bm{x}})=\frac{1}{2}\bm{H}(\bm{x})^T\bm{H}(\bm{x}),
\end{equation}
Starting with an initial guess $\bm{x}^{[0]}$, the Gauss-Newton method proceeds by the iterations
\begin{equation}\label{gaussnewton}
\bm{x}^{[j+1]}=\bm{x}^{[j]}-(\bm{J}^\mathsf{T} \bm{J})^{-1}\bm{J}^\mathsf{T} \bm{H}\mid_{\bm{x}=\bm{x}^{[j]}},
\end{equation}
where $\bm{x}^{[j]}$ ($j\geq1$) is the $j$-th iterative output and $\bm{J}$ is the Jacobian matrix of $\bm{H}$, i.e.
\begin{equation}\label{jacobian}
\bm{J}=\left[\frac{\partial H_i}{\partial x_j}\right]_{i=1,\ldots,2^{N+1};j=1,\ldots,N+2+N(N+1)/2}
\end{equation}
This iterate process makes the  objective function $S({\bm{x}})$ decay to zero. In the numerical experiments, if $\bm{J}^\mathsf{T} \bm{J}$ is near singular, change it to $\bm{J}^\mathsf{T} \bm{J}+10^{-6} \bm{E}$ to modify the singularity, where $\bm{E}$ is the unit matrix.

The key  of the procedure is the choice of initial guess $\bm{x}^{[0]}$. We suggest the following guidance to determine the initial guess.  For given $k_j$, solve the initial guess  $\omega_j^{[0]}$ and $l_j^{[0]}$ from the equations
\begin{eqnarray}
&&F_1(i\omega_j^{[0]},il_j^{[0]},ik_j,\cdots,c_1^{[0]})=0,\label{qd1}\\
&&F_2(i\omega_j^{[0]},il_j^{[0]},ik_j,\cdots,c_2^{[0]})=0,\label{qd2}
\end{eqnarray}
where the initial guess of $c_1^{[0]}$,$c_2^{[0]}$ are generally taken to be $\pm1$.
In fact, if the initial guess $\bm{x}^{[0]}$ satisfies  Eqs.~\eqref{qd1}-\eqref{qd2},  the objective function  $S({\bm{x}})$  will have a smaller initial value.

\section{Numerical results}
\label{section4}
In this section, we use the numerical scheme to calculate $1$-, $2$- and $3$-periodic wave solutions of the coupled Ramani Eqs.~\eqref{cramani1}-\eqref{cramani2}.
This system was first proposed in Ref.~\cite{cramani1},  and  its $N$-solitons was known to be expressed by Pfaffians\cite{cramani2}.  Some other properties and generalizations can be found in Refs.~\cite{cramani3,cramani4,cramani5,cramani6}. As far as we know, there is no results about the periodic waves of this equation.
It is worth mentioning that when  $v=0$, the coupled Ramani equation reduces to the following Ramani equation\cite{ramani1}.
\begin{eqnarray}
&&u_{xxxxxx}+15u_{xx}u_{xxx}+15u_xu_{xxxx}+45u_x^2u_{xx}\nonumber\\
&&\quad-5(u_{xxxt}+3u_{xx}u_t+3u_xu_{xt})-5u_{tt}=0.\label{ramani}
\end{eqnarray}

Note that there are two constants $u_0$ and $v_0$ in the variable transformation \eqref{vtran1} and $u_0$ makes no difference to the bilinear equations while $v_0$ does. Thus the numerical experiments will be carried out with $v_0=0$ and $v_0=1$ respectively.  When plotting the profile of $u$ and $v$, we will take $u_0=0$, $\eta_j^{0}=0$ and  $z=0$ without loss of generality.

All computations are carried out in Matlab R2013b on a computer with a 2.83 GHz CPU and 8 GB main memory.  The termination condition for stopping the numerical iterative is $||{\bm{x}^{[j+1]}}-{\bm{x}^{[j]}}||_2<10^{-14}$ and $||{\bm{H}}||_2<10^{-14}$ where $||\large{\cdot}||_2$ means $2$-norm.

\subsection{$1$-periodic waves}
In the case of $N=1$, according to Theorem 2.1,  the problem is for given $k_1$ and $\tau_{11}$, to solve $\omega_1$, $l_1$, $c_1$ and $c_2$ from a nonlinear algebraic system of 4 equations. Note that the coupled Ramani equation \eqref{cramani1}-\eqref{cramani2} are linear in $c_1$ and $c_2$, and  include  terms $D_t^2$ and $D_zD_t$. Thus after a tedious calculation, the nonlinear algebraic system reduces to a cubic equation of $\omega_1$. Therefore, although very tediously but possibly, we are able to write out the exact solutions.

\begin{figure*}[!ht]
   \centering
     \includegraphics[totalheight=11pc]{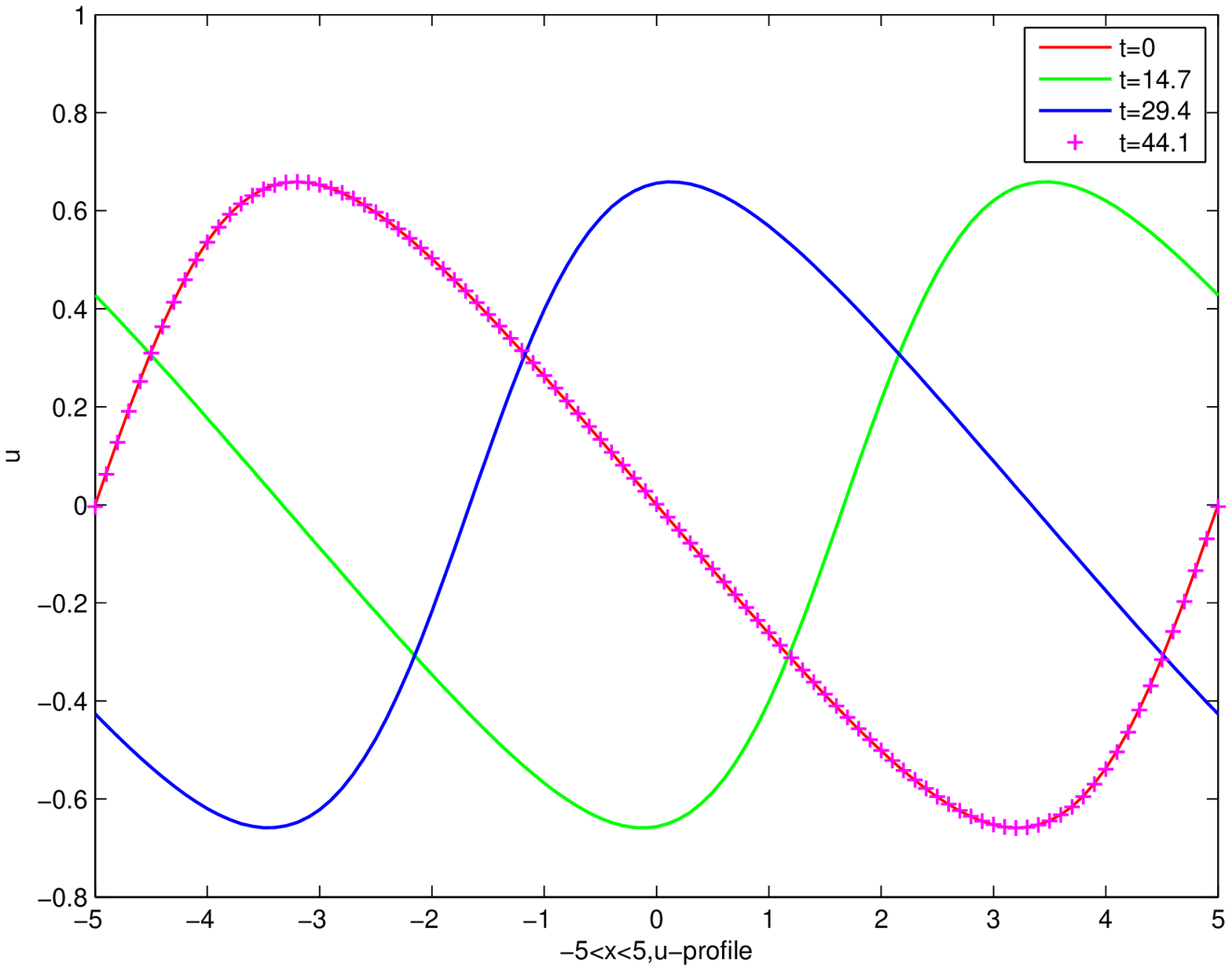}
     \includegraphics[totalheight=11pc]{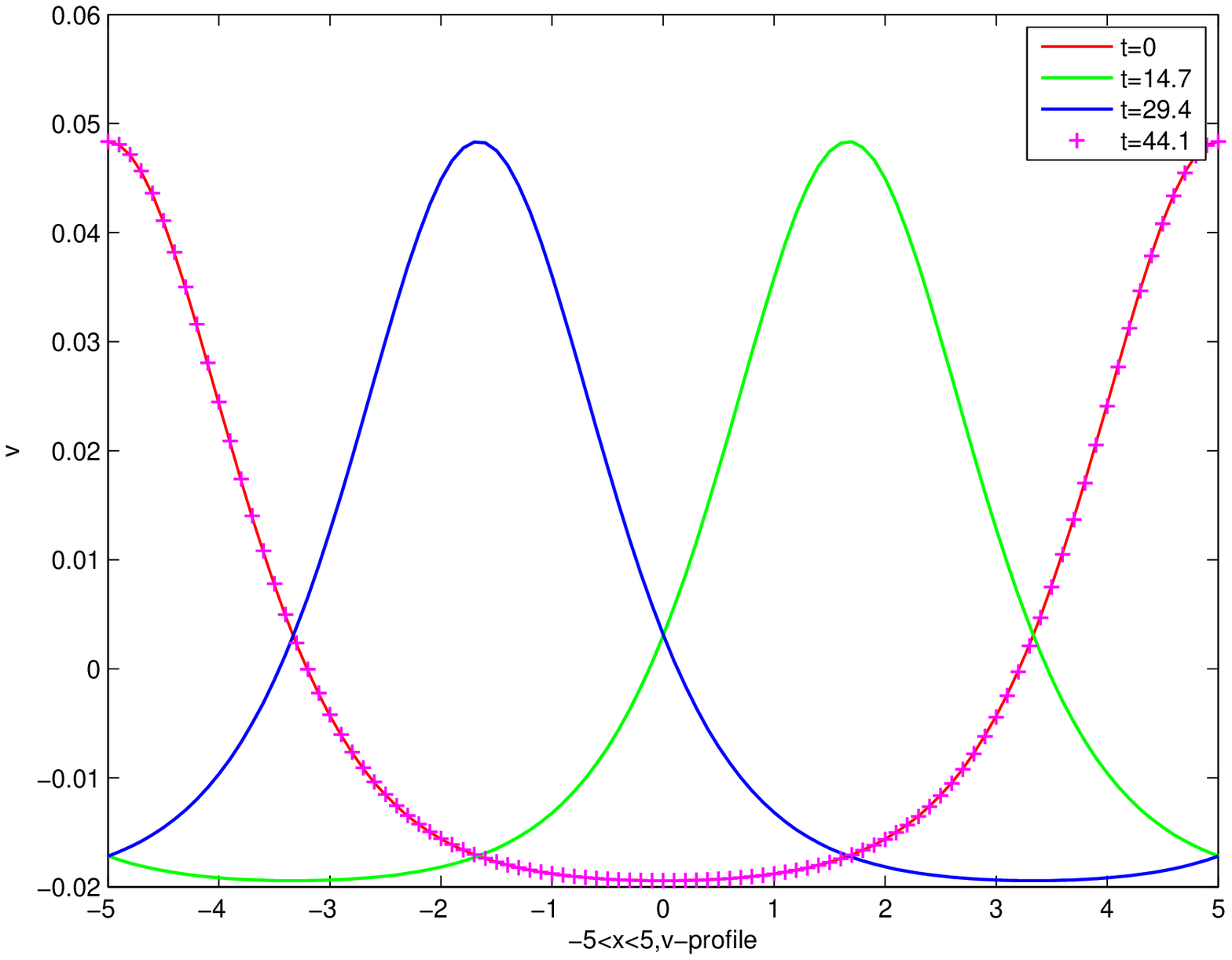}
         \caption{Profile of $u$ and $v$ of the first example of Table~\ref{table11}.  Figure left: $u$-profile;  Figure right: $v$-profile}
      \label{uv1}
\end{figure*}

Here, instead of giving the exact expressions, we will present some $1$-periodic wave solutions numerically by using the numerical scheme given in Sect.~3. When $N=1$, the Jacobian matrix $\bm{J}$ is $4\times 4$ and the Gauss-Newton iteration \eqref{gaussnewton} reduces to the Newton iteration
\begin{equation}
\bm{x}^{[j+1]}=\bm{x}^{[j]}-\bm{J}^{-1}\bm{H}\mid_{\bm{x}=\bm{x}^{[j]}}.
\end{equation}
The numerical experiments are successful and the errors of $||{\bm{H}}||_2=0$ hold within $\sim10^{-15}$. See Tables~\ref{table11}-\ref{table12} for several detailed examples. Fig.~\ref{uv1} shows the profiles of $u$ and $v$ of the first example in Table~\ref{table11}.  This $1$-periodic wave solution is periodic both in time and space. Actually, its spatial and temporal periods   are $\frac{2\pi}{k_1}=10$ and $\frac{2\pi}{l_1}=44.1235$ respectively.  As shown in the Fig.~\ref{uv1}, the profiles at $t=0$ and $t=44.1$  almost coincide for both $u$ and $v$.

 In the case of $v_0=0$, the numerical experiments may give some solutions
of $l_1=0$(see the second example in Table~\ref{table11}) which means that the Riemann's $\theta$-function is independent on $z$. With the variable transformation \eqref{vtran1},  we have $v=0$. Therefore, in this case,  the solution $u$ reduces to the solution of the Ramani equation \eqref{ramani}.  The same goes in the numerical experiments of $2$-periodic and $3$-periodic waves which will be given below.

 \begin{table}[!ht]
\center
\begin{tabular}{|c|c|c|c|c|c|c|}
\hline
  $k_1$                               &   $\tau_{11}$                     & $c_1^{[0]}$       &  $c_2^{[0]}$\\
\hline
  $1\times\frac{2\pi}{10}$   &    $0.46\times 2\pi$           &  $1$                  &  $1$\\
\hline
  $\omega_1$                    &  $l_1$                                &  $c_1$              &  $c_2$   \\
\hline
  $0.1424$                        &  $0.0921$                          &  $0.8494$        &   $0.0419$\\
  \hline
  \hline
  $k_1$                               &   $\tau_{11}$                     & $c_1^{[0]}$       &  $c_2^{[0]}$\\
\hline
  $1\times\frac{2\pi}{10}$   &    $1.86\times 2\pi$           &  $1$                  &  $1$\\
\hline
  $\omega_1$                    &  $l_1$                                &  $c_1$              &  $c_2$   \\
\hline
  $-0.0423$                        &  $0$                          &  $0.00005$        &   $0$\\  \hline
\end{tabular}
\caption{$1$-periodic waves: examples with $v_0=0$ }
\label{table11}
\end{table}

 \begin{table}[!ht]
\center
\begin{tabular}{|c|c|c|c|c|c|c|}
\hline
  $k_1$                               &   $\tau_{11}$                     & $c_1^{[0]}$       &  $c_2^{[0]}$\\
\hline
  $1\times\frac{2\pi}{10}$   &    $0.46\times 2\pi$           &  $1$                  &  $1$\\
\hline
  $\omega_1$                    &  $l_1$                                &  $c_1$              &  $c_2$   \\
\hline
  $1.3800$                        &  $1.9139$                          &  $3.6650$        &   $0.8708$\\
  \hline
  \hline
  $k_1$                               &   $\tau_{11}$                     & $c_1^{[0]}$       &  $c_2^{[0]}$\\
\hline
  $1\times\frac{2\pi}{10}$   &    $1.86\times 2\pi$           &  $1$                  &  $1$\\
\hline
  $\omega_1$                    &  $l_1$                                &  $c_1$              &  $c_2$   \\
\hline
  $1.4071$                        &  $1.4312$                          &  $0.0004$        &   $0.0001$\\  \hline
\end{tabular}
\caption{$1$-periodic waves: examples with $v_0=1$}
\label{table12}
\end{table}

\subsection{$2$-periodic waves}
\begin{figure*}[!ht]
   \centering
     \includegraphics[totalheight=12pc]{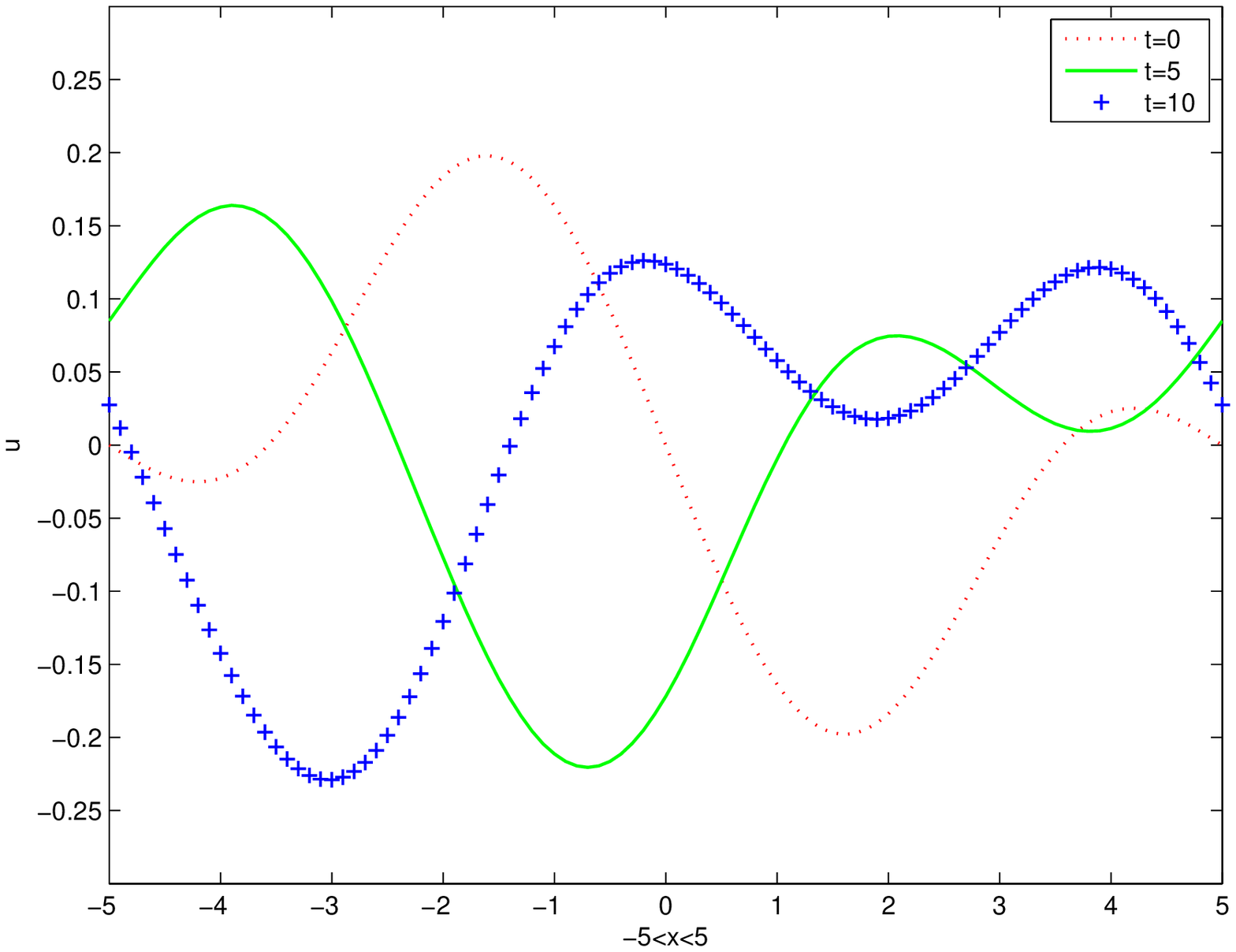}
     \includegraphics[totalheight=12pc]{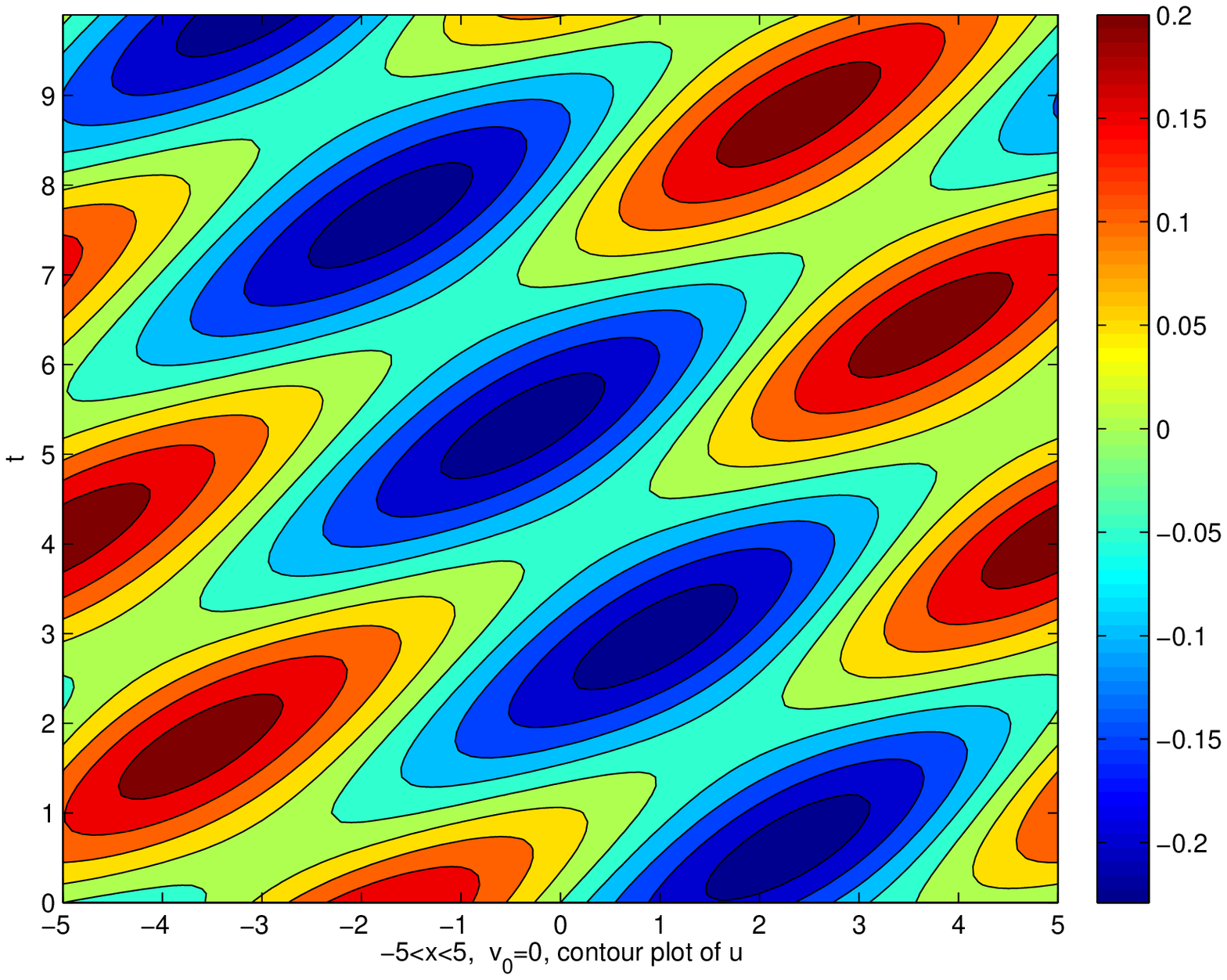}\\
     \includegraphics[totalheight=12pc]{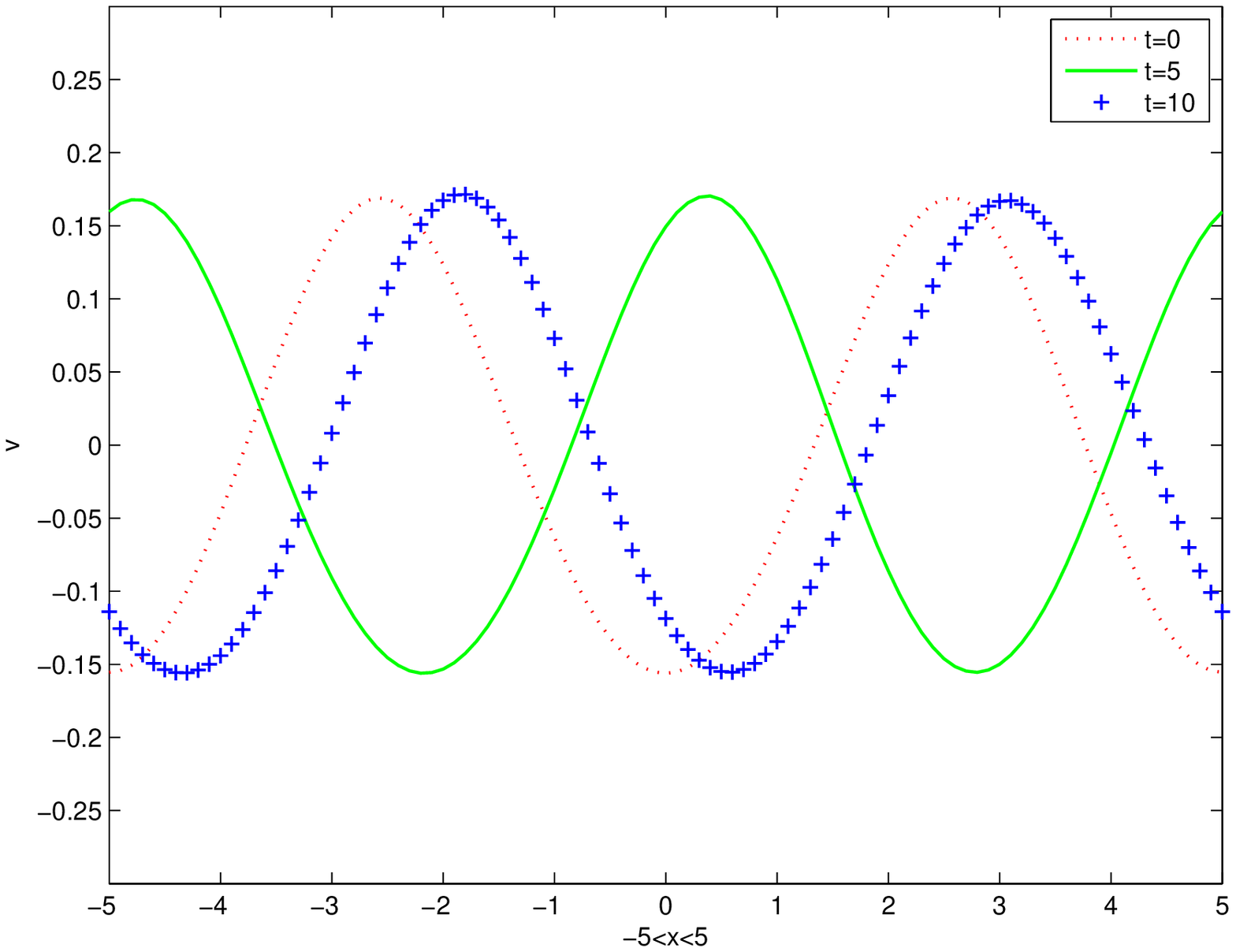}
     \includegraphics[totalheight=12pc]{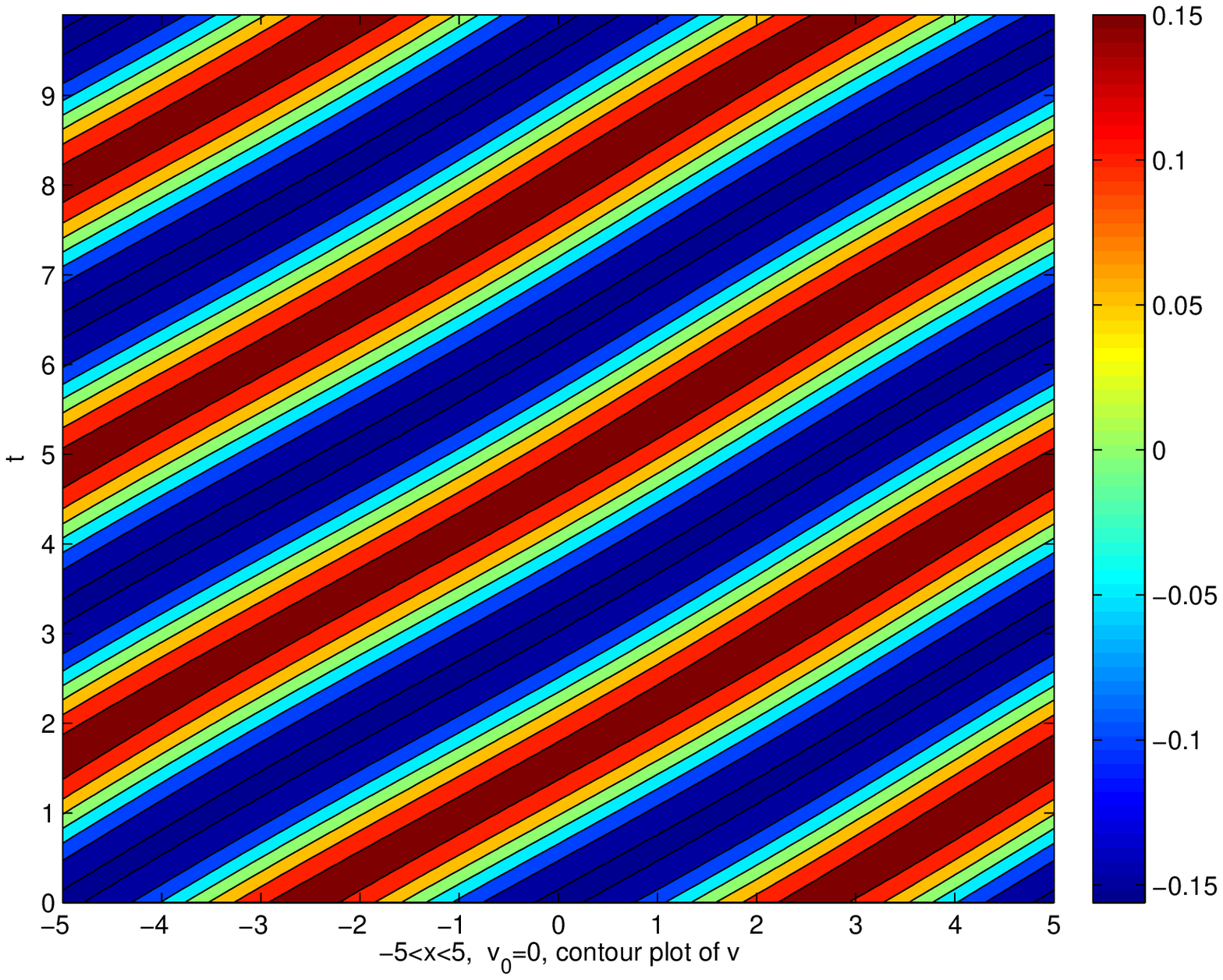}
         \caption{The first example of Table~\ref{table21}.  Top left: $u$-profile;  Top right: contour plot of $u$; Below left: $v$-profile;  Below right: contour plot of $v$.}
      \label{uv21}
\end{figure*}

\begin{figure*}[!ht]
   \centering
     \includegraphics[totalheight=12pc]{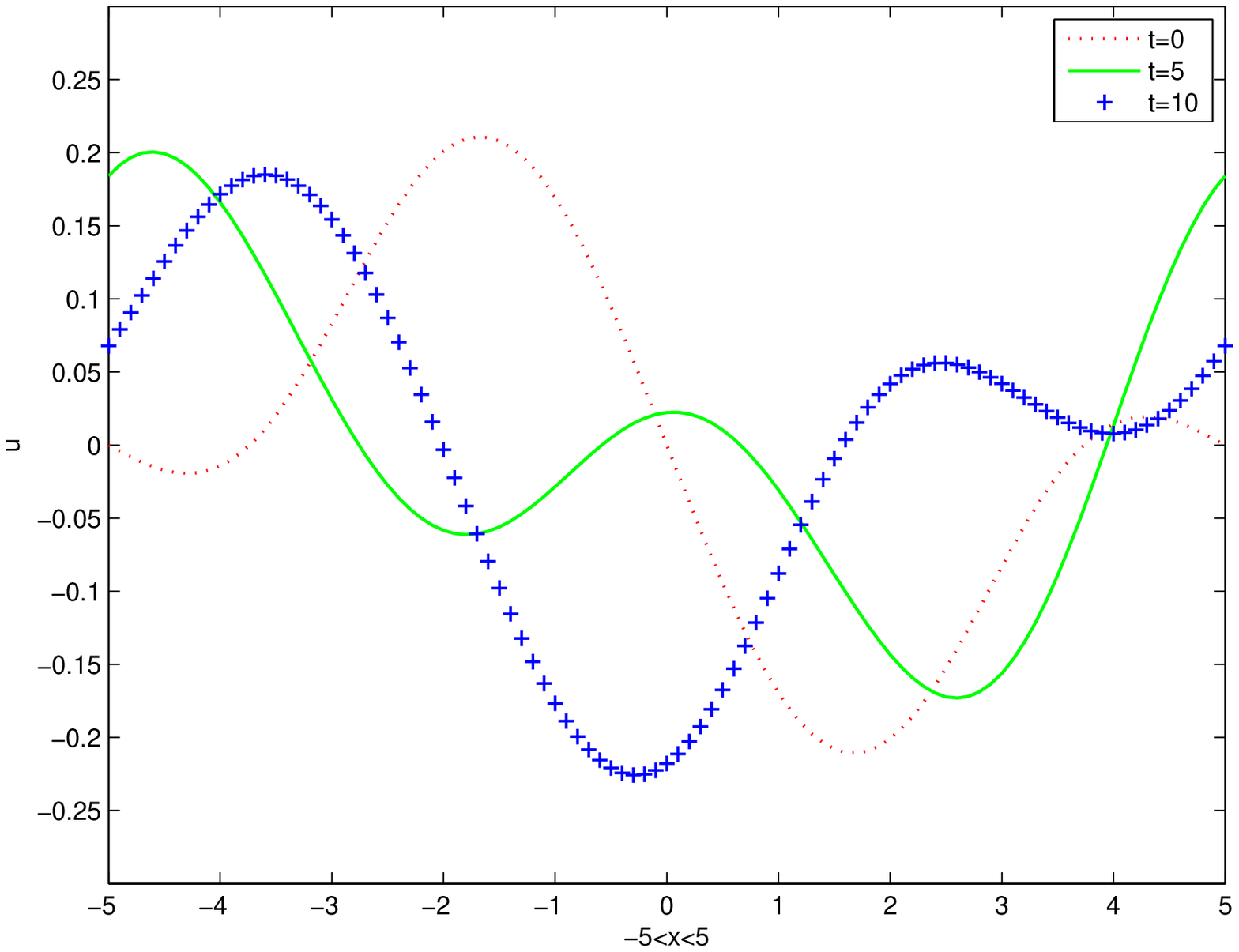}
     \includegraphics[totalheight=12pc]{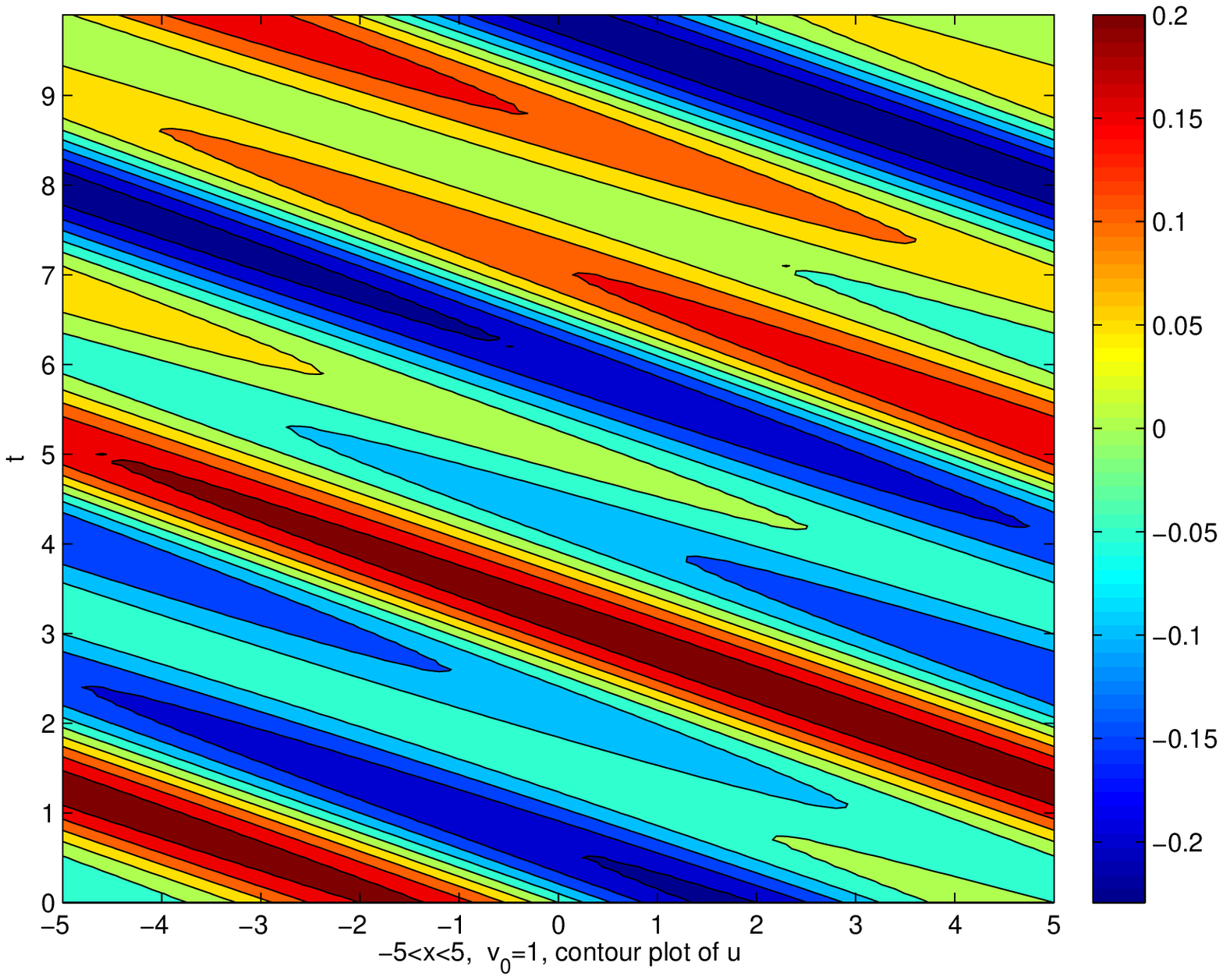}\\
     \includegraphics[totalheight=12pc]{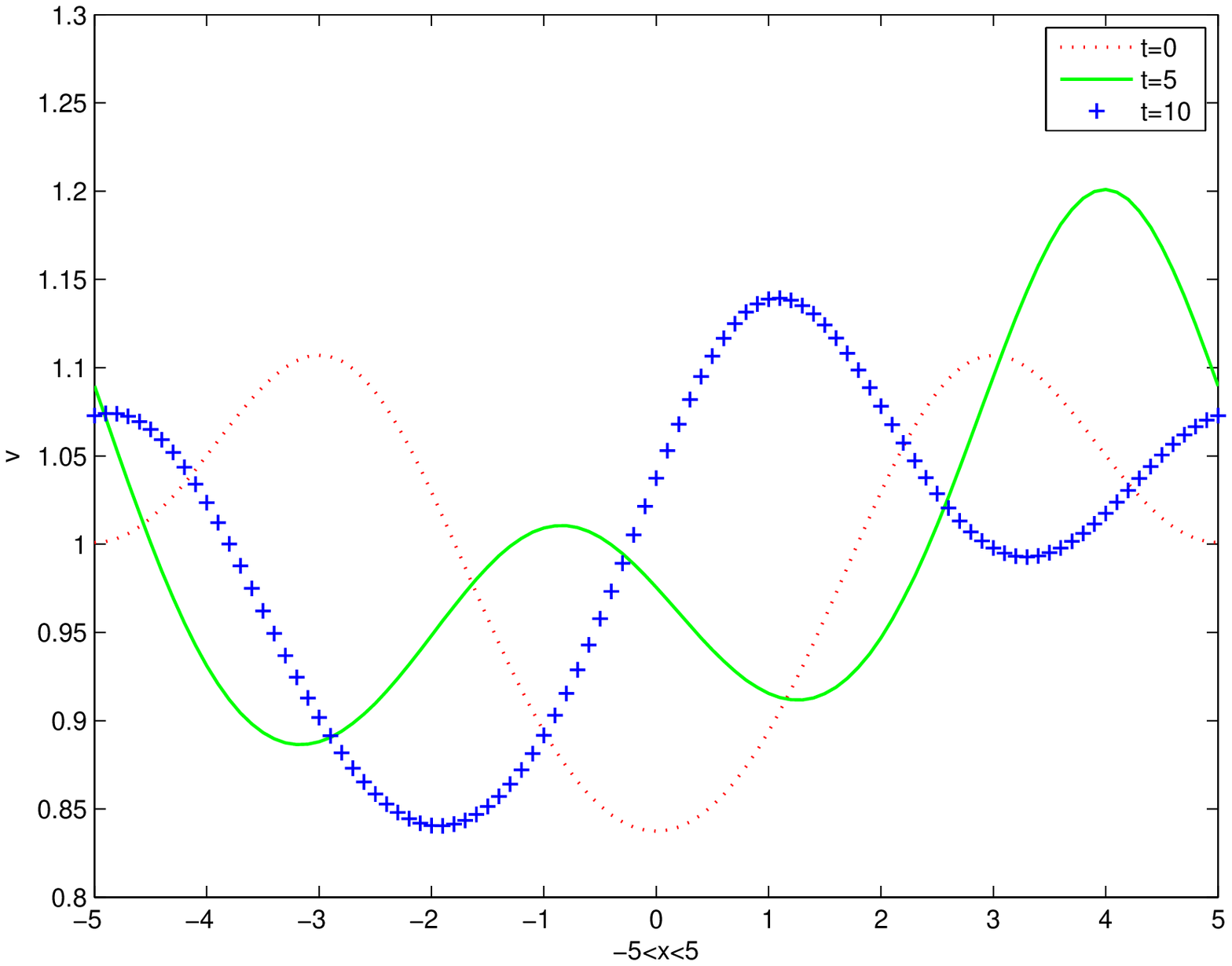}
     \includegraphics[totalheight=12pc]{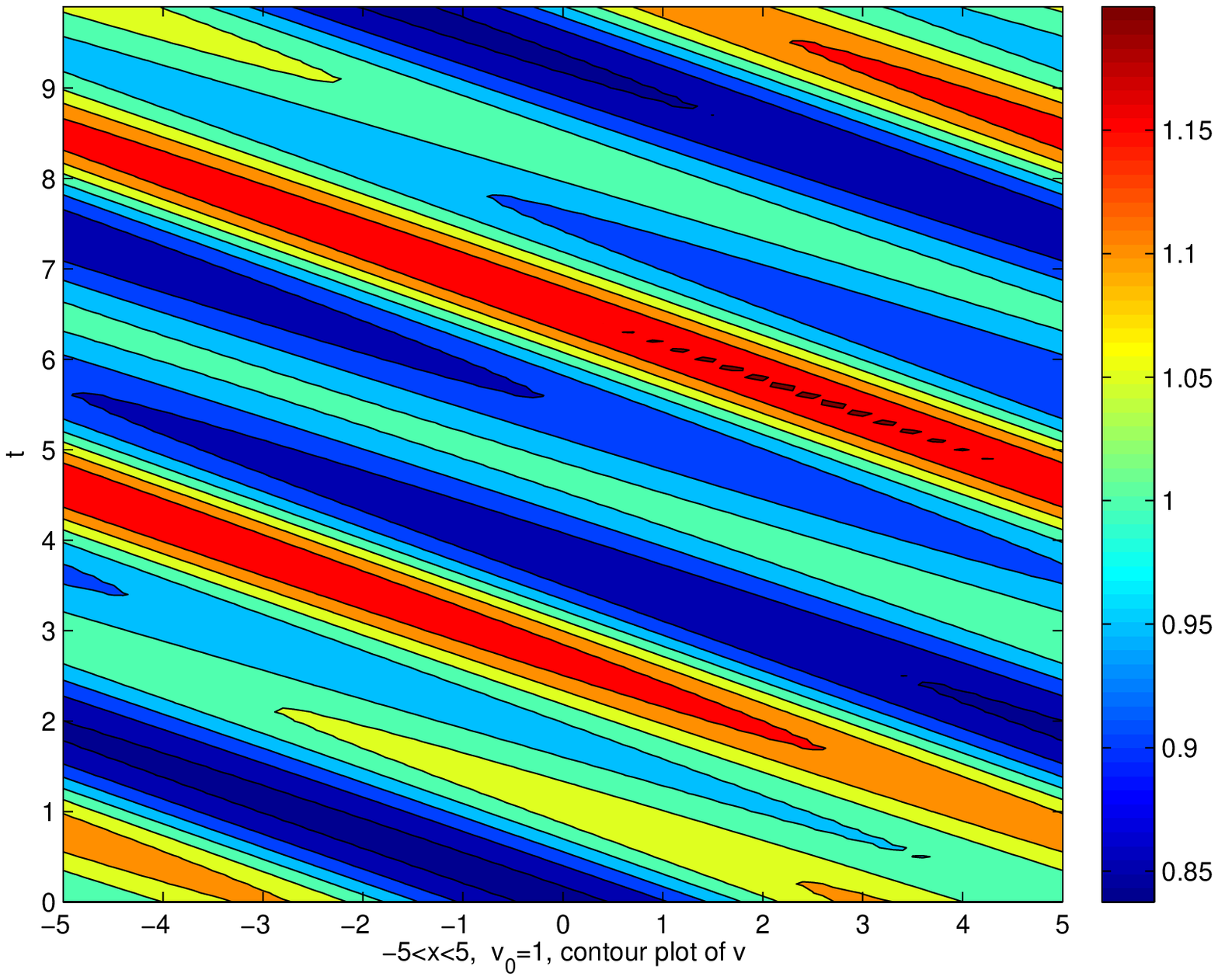}
         \caption{The first example of Table~\ref{table22}.  Top left: $u$-profile;  Top right: contour plot of $u$; Below left: $v$-profile;  Below right: contour plot of $v$.}
      \label{uv22}
\end{figure*}

In this case, the nonlinear algebraic system \eqref{con2}-\eqref{con3} is an over-determined system of 8 equations with 7 unknowns.  In our numerical experiments, the errors of $||{\bm{H}}||_2=0$  also hold within $\sim10^{-15}$. See the detailed examples in Tables~\ref{table21}-\ref{table22} and Figs.~\ref{uv21}-\ref{uv22}.  These waves are periodic in space but only quasi-periodic in time.  We also give an example of $l_j=0(j=1,2)$(see the last example in Table~\ref{table21}).
 \begin{table*}[!ht]\tiny
\center
\begin{tabular}{|c|c|c|c|c|c|c|c|c|c|c|c|c|c|c|c|c|}
\hline
$k_1$   &$k_2$    &   $\tau_{11}$   &  $\tau_{22}$     & $c_1^{[0]}$      &  $c_2^{[0]}$   &$\omega_1$     & $\omega_2$   &  $l_1$      &  $l_2$   & $\tau_{12}$    &  $c_1$     &  $c_2$    \\
\hline
  $1\times\frac{2\pi}{10}$   &  $2\times\frac{2\pi}{10}$   &  $0.96\times 2\pi$    &  $1.23\times 2\pi$       &  $1$           &  $1$ &
    $0.3556$   &  $-1.9620$   &     $0.0313$     &       $3.0793$   &     $0.9060$   &   $0.0460$   &  $0.0651$    \\
  \hline
\hline
  $k_1$                   &$k_2$        &   $\tau_{11}$       &  $\tau_{22}$              & $c_1^{[0]}$       &  $c_2^{[0]}$   & $\omega_1$       &$\omega_2$             &  $l_1$         &  $l_2$           & $\tau_{12}$            &  $c_1$              &  $c_2$ \\
\hline
  $1\times\frac{2\pi}{10}$   &  $2\times\frac{2\pi}{10}$   &  $0.46\times 2\pi$    &  $1.03\times 2\pi$       &  $1$       &  $1$ &
    $0.3724$   &    $-1.8403$   &  $0.2439$   &  $2.9825$   &  $1.0610$   &    $1.4268$   &  $0.3234$   \\
  \hline
  \hline
  $k_1$                   &$k_2$        &   $\tau_{11}$       &  $\tau_{22}$              & $c_1^{[0]}$       &  $c_2^{[0]}$   & $\omega_1$       &$\omega_2$             &  $l_1$         &  $l_2$           & $\tau_{12}$            &  $c_1$              &  $c_2$  \\
\hline
  $1\times\frac{2\pi}{10}$   &  $2\times\frac{2\pi}{10}$   &  $0.52\times 2\pi$    &  $1.13\times 2\pi$       &  $1$       &  $-1$ &
    $-0.2612$   &    $-0.8778$   &  $0$   &  $0$   &  $-0.5938$   &    $0.3925$   &  $0$   \\
  \hline
\end{tabular}
\caption{$2$-periodic waves: examples with $v_0=0$}
\label{table21}
\end{table*}
\begin{table*}[!ht]\tiny
\center
\begin{tabular}{|c|c|c|c|c|c|c|c|c|c|c|c|c|c|c|c|c|}
\hline
  $k_1$                   &$k_2$        &   $\tau_{11}$       &  $\tau_{22}$              & $c_1^{[0]}$       &  $c_2^{[0]}$   &  $\omega_1$       &$\omega_2$             &  $l_1$         &  $l_2$           & $\tau_{12}$            &  $c_1$              &  $c_2$\\
\hline
  $1\times\frac{2\pi}{10}$   &  $2\times\frac{2\pi}{10}$   &  $0.96\times 2\pi$    &  $1.23\times 2\pi$       &  $1$       &  $1$ &
        $1.4078$  &  $3.3761$   &   $1.4515$   &    $1.7779$   &  $2.2216$   &   $0.6755$   &  $0.0575$   \\
  \hline
\hline
  $k_1$                   &$k_2$        &   $\tau_{11}$       &  $\tau_{22}$              & $c_1^{[0]}$       &  $c_2^{[0]}$   & $\omega_1$       &$\omega_2$             &  $l_1$         &  $l_2$           & $\tau_{12}$            &  $c_1$              &  $c_2$\\
\hline
  $1\times\frac{2\pi}{10}$   &  $2\times\frac{2\pi}{10}$   &  $0.46\times 2\pi$    &  $1.03\times 2\pi$       &  $1$           &  $1$ &
    $1.3897$   &  $3.3863$   &     $1.9282$     &       $1.9582$   &     $1.9332$   &   $5.1707$   &  $0.9571$    \\
  \hline
    \hline
  $k_1$                   &$k_2$        &   $\tau_{11}$       &  $\tau_{22}$              & $c_1^{[0]}$       &  $c_2^{[0]}$   & $\omega_1$       &$\omega_2$             &  $l_1$         &  $l_2$           & $\tau_{12}$            &  $c_1$              &  $c_2$\\
\hline
  $1\times\frac{2\pi}{10}$   &  $2\times\frac{2\pi}{10}$   &  $0.52\times 2\pi$    &  $1.13\times 2\pi$       &  $1$       &  $-1$ &
        $1.3843$   &       $3.3920$   &       $1.7341$   &       $1.8698$   &       $2.0174$  &     $3.1582$   &  $0.5405$   \\
  \hline
\end{tabular}
\caption{$2$-periodic waves: examples with $v_0=1$}
\label{table22}
\end{table*}

\subsection{$3$-periodic waves}
\begin{figure*}[!ht]
   \centering
     \includegraphics[totalheight=12pc]{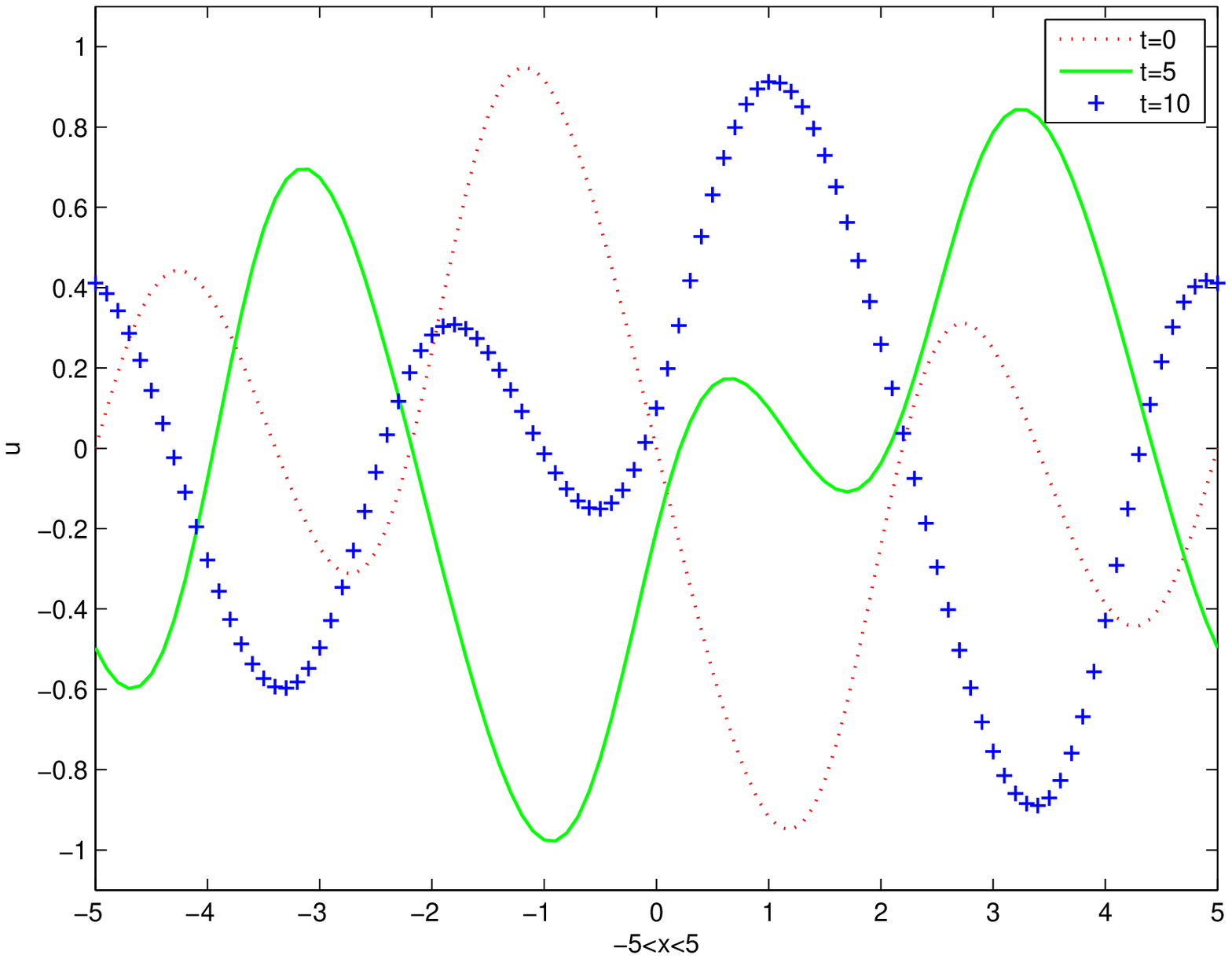}
     \includegraphics[totalheight=12pc]{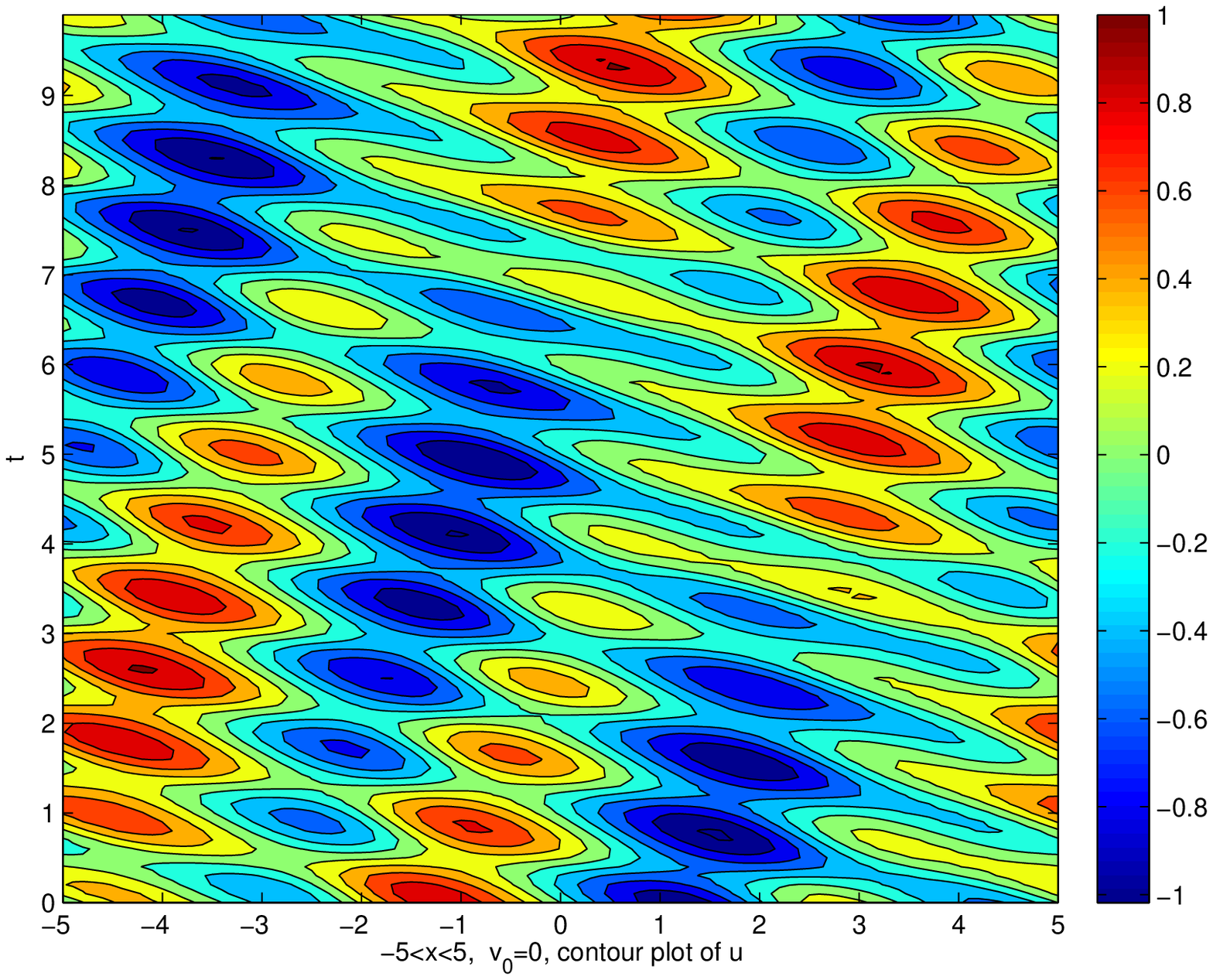}\\
     \includegraphics[totalheight=12pc]{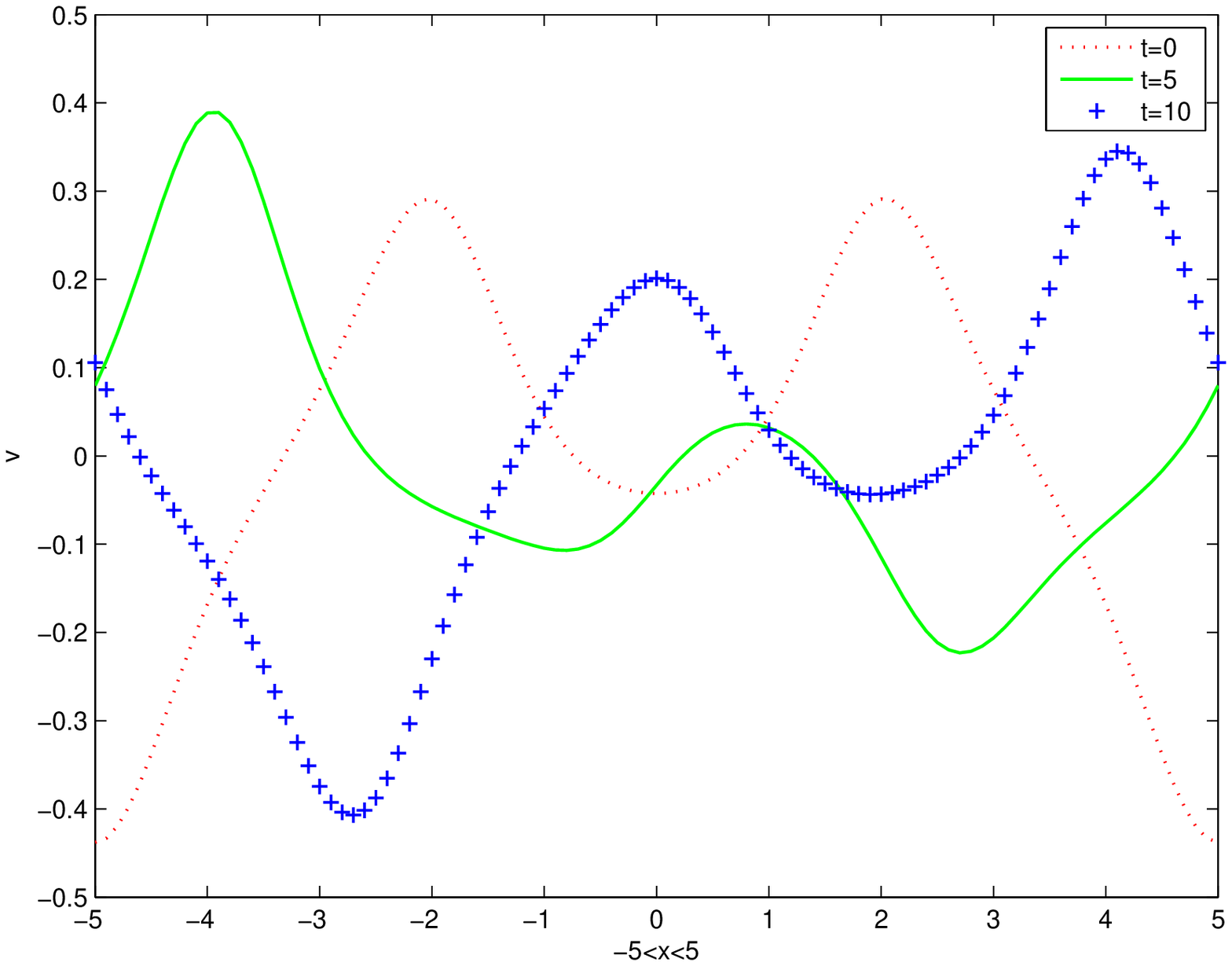}
     \includegraphics[totalheight=12pc]{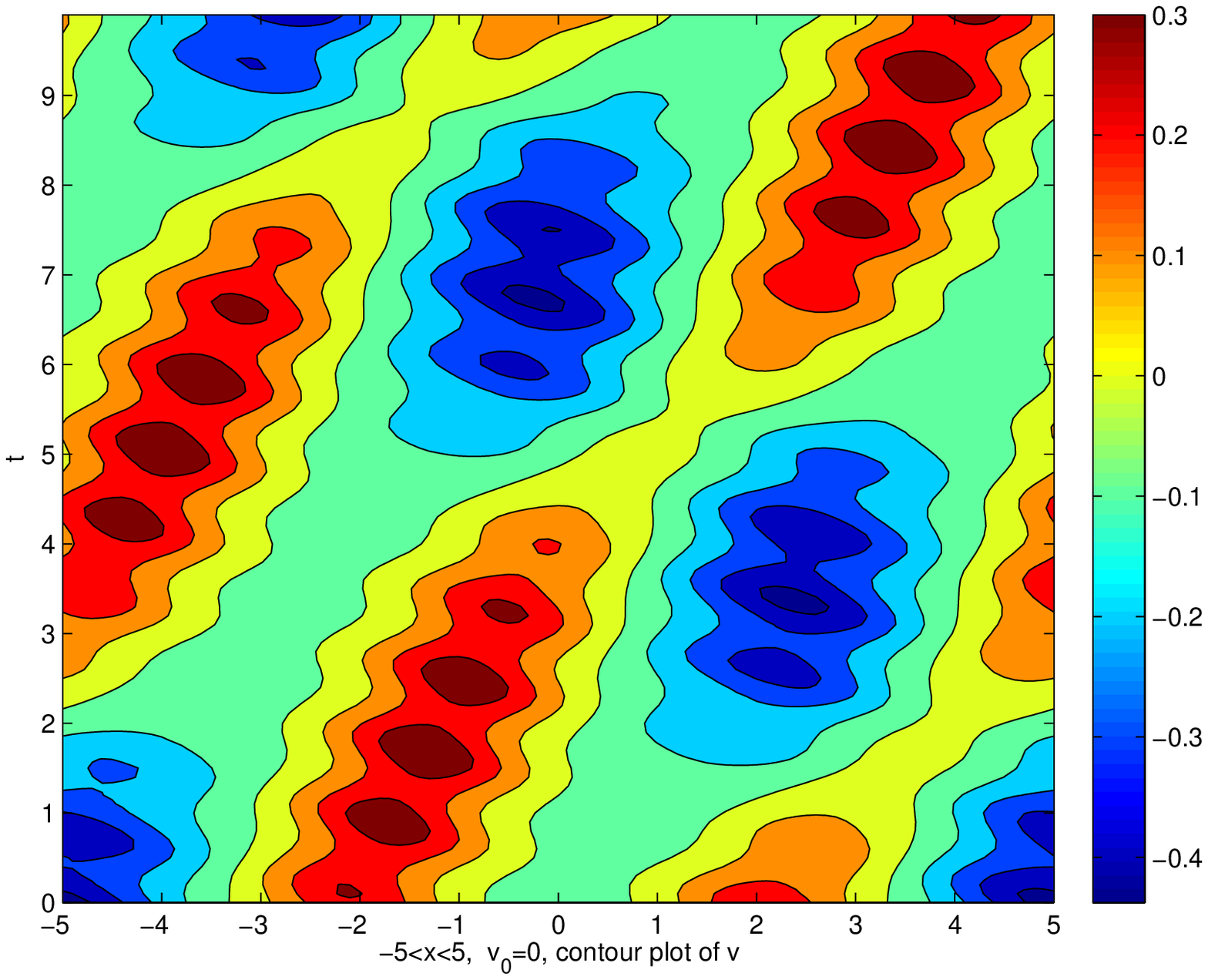}
         \caption{The first example of Table~\ref{table31}.  Top left: $u$-profile;  Top right: contour plot of $u$; Below left: $v$-profile;  Below right: contour plot of $v$.}
      \label{uv31}
\end{figure*}

\begin{figure*}[!ht]
   \centering
     \includegraphics[totalheight=12pc]{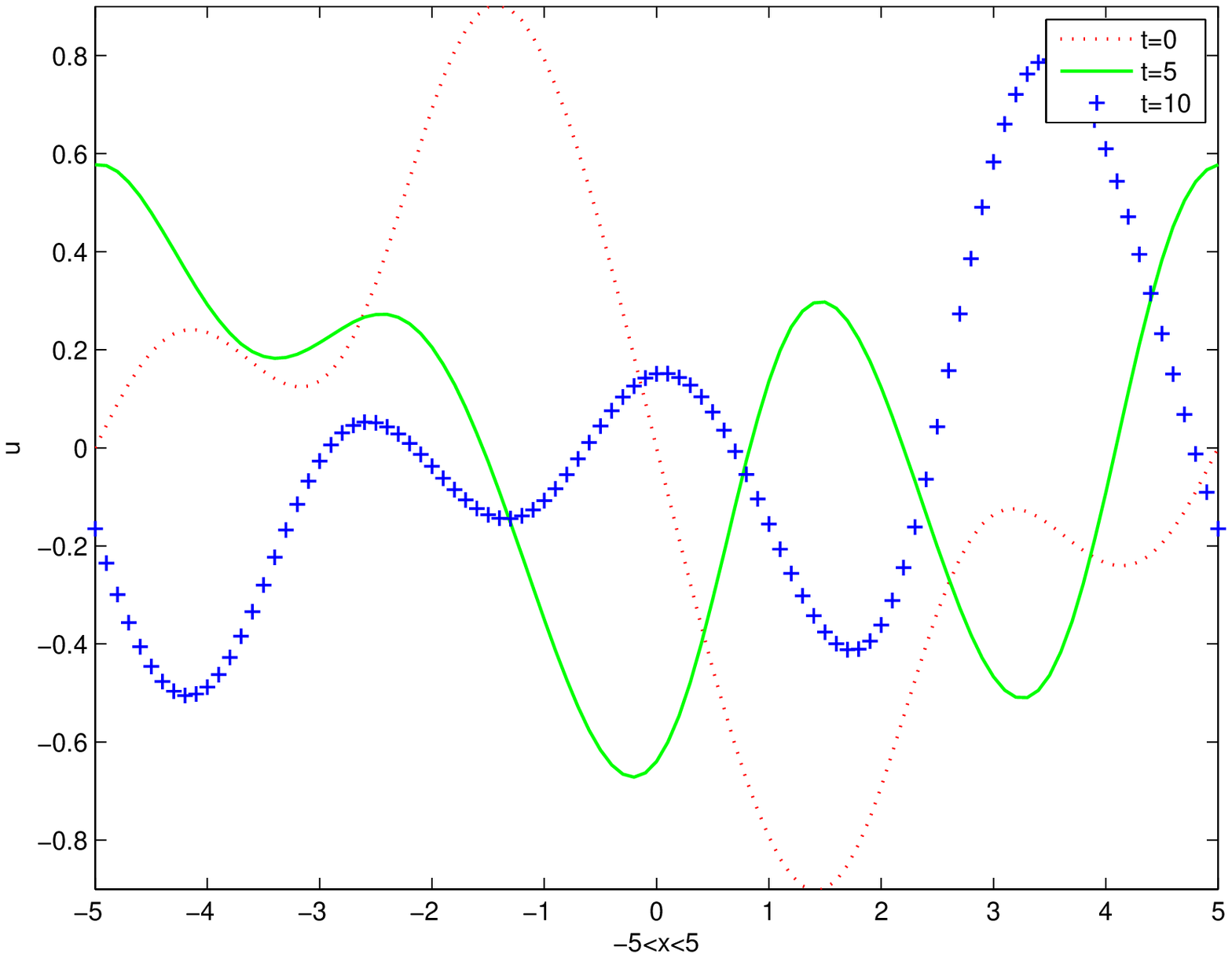}
     \includegraphics[totalheight=12pc]{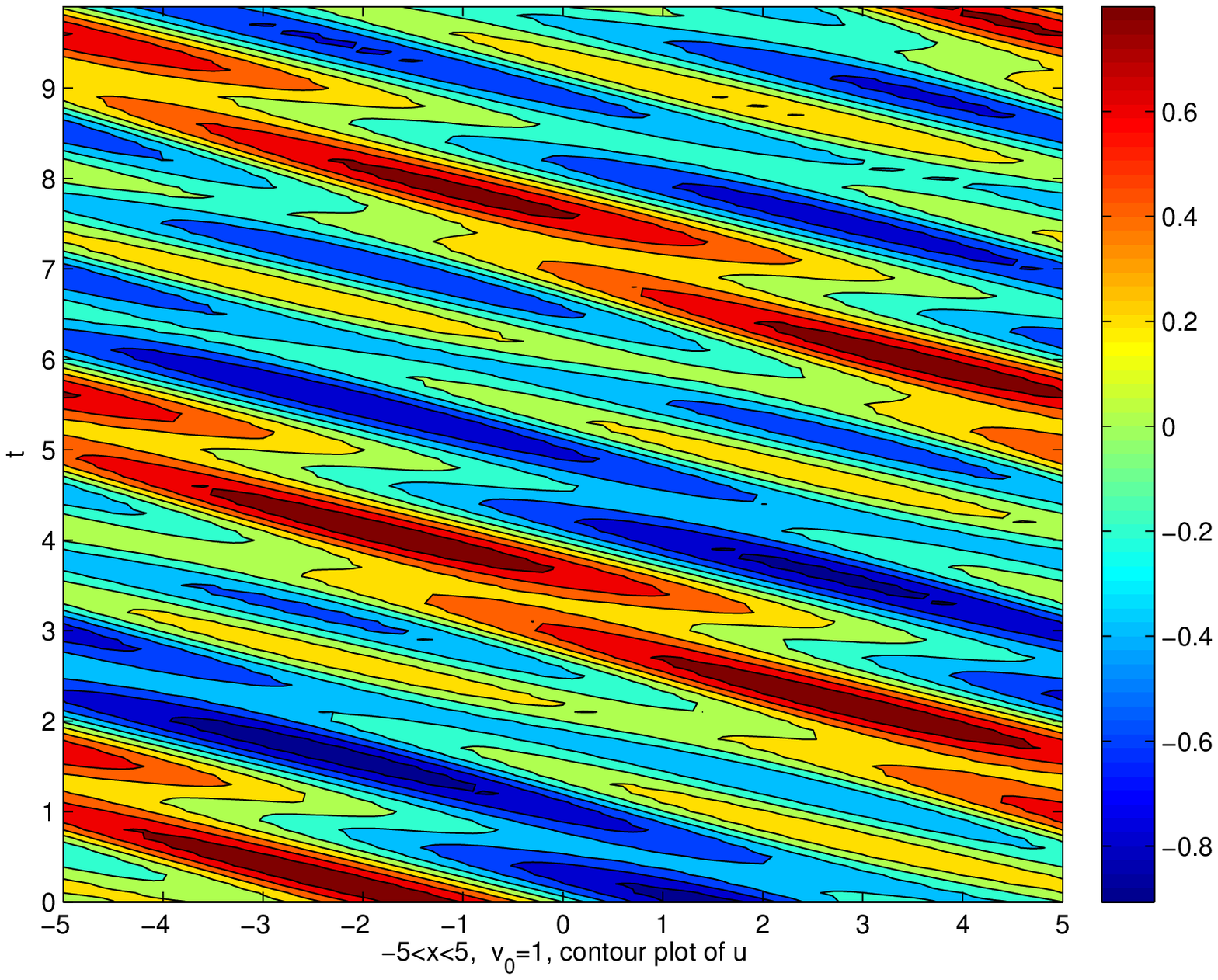}\\
     \includegraphics[totalheight=12pc]{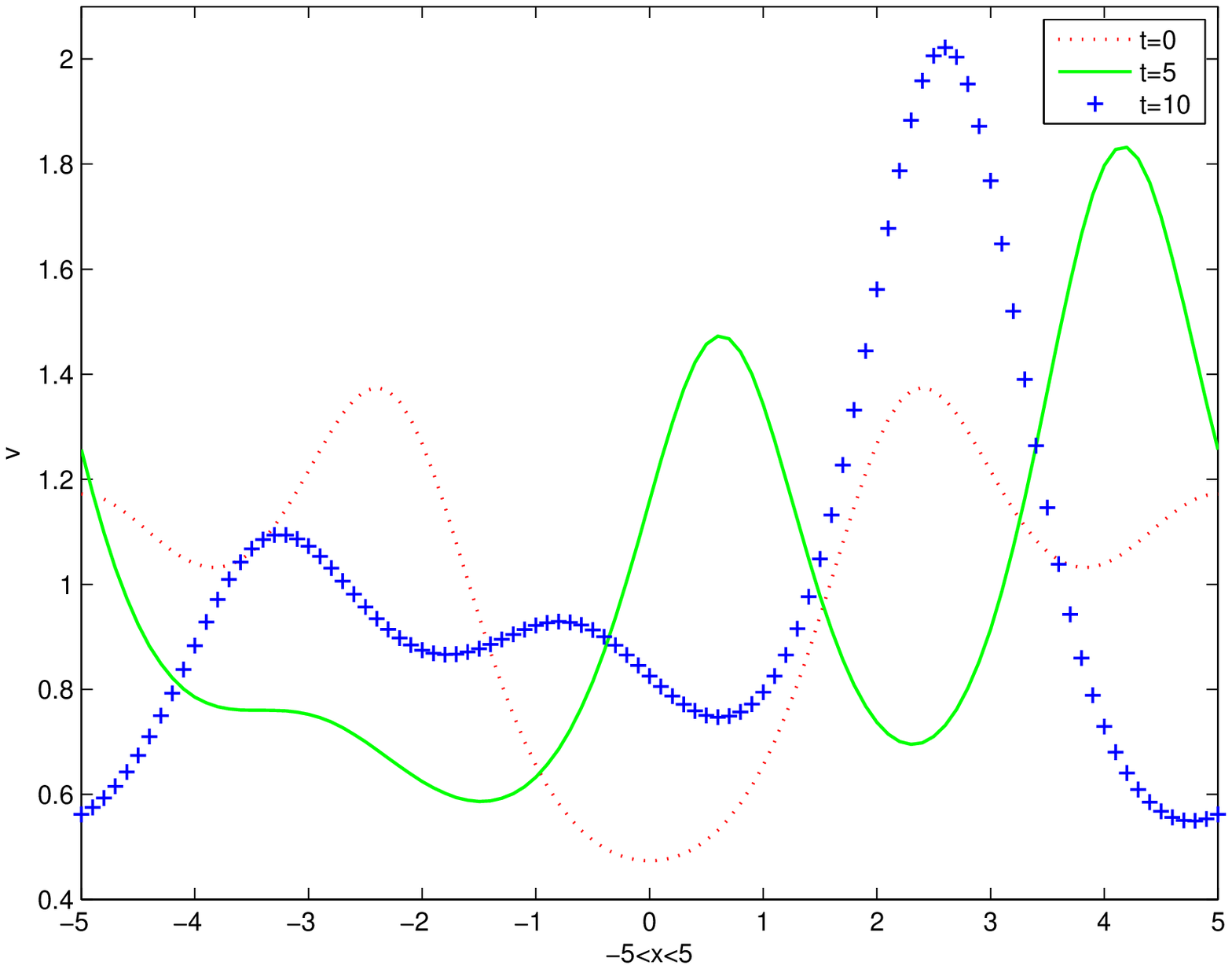}
     \includegraphics[totalheight=12pc]{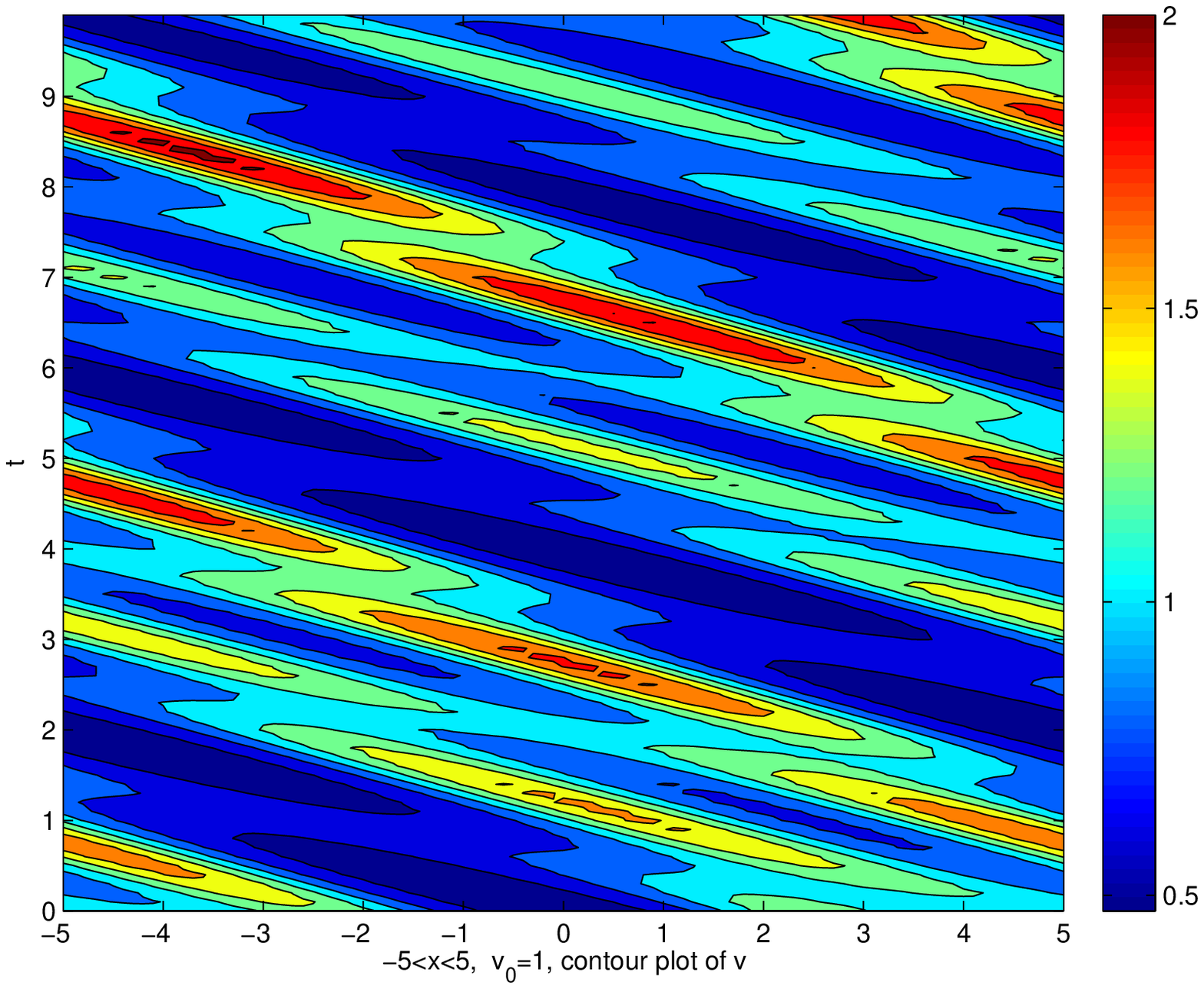}
         \caption{The first example of Table~\ref{table32}.  Top left: $u$-profile;  Top right: contour plot of $u$; Below left: $v$-profile;  Below right: contour plot of $v$.}
      \label{uv32}
\end{figure*}

In the case of $3$-periodic waves,  the problem is a nonlinear system of 16 equations with 11 unknowns.
Some detailed numerical examples are given  in Tables~\ref{table31}-\ref{table32}. As shown in Figs.~\ref{uv31}-\ref{uv32}, the $3$-periodic wave represents three waves interact with each other repeatedly.

In some cases, the numerical experiments will produce results with $l_j=0$(see the third example in Table~\ref{table31}).  As we stated before, this kind of solutions will reduce to the solutions of the Ramani equation \eqref{ramani}.  However, as the iterative output ${\bm{x}}^{[j]}$ tends to the exact solution, the corresponding matrix $\bm{J}^\mathsf{T} \bm{J}$ will be near-singular  which will result in accuracy degradation.  Actually, in the third example of Table~\ref{table31}, the error of $||{\bm{H}}||_2=0$ only holds within $\sim10^{-9}$  while in the other examples, the errors hold within $\sim10^{-14}$.

\begin{table*}[!ht]\tiny
\center
\begin{tabular}{|c|c|c|c|c|c|c|c|c|c|c|}
\hline
  $k_1$   &$k_2$    & $k_3$   &$\tau_{11}$     & $\tau_{22}$      &$\tau_{33}$     & $c_1^{[0]}$     & $c_2^{[0]}$   & & &\\
\hline
  $1\times\frac{2\pi}{10}$   &  $2\times\frac{2\pi}{10}$  &  $3\times\frac{2\pi}{10}$  &  $0.67\times 2\pi$    &  $0.86\times 2\pi$  &  $1.02\times 2\pi$       &  $1$       &  $1$ &&&\\
\hline
  $\omega_1$       &$\omega_2$    &  $\omega_3$        &  $l_1$         &  $l_2$      &  $l_3$     & $\tau_{12}$     & $\tau_{13}$    & $\tau_{23}$           &  $c_1$              &  $c_2$   \\
\hline
   $0.4685$   &$-0.8643$   & $7.0815$  &$-0.9501$  &$1.0718$  &$0.0183$  & $-1.4992$  &$1.0605$  &$1.6167$  &$24.5355$&  $0.1485$   \\
  \hline
  \hline
  $k_1$   &$k_2$    & $k_3$   &$\tau_{11}$     & $\tau_{22}$      &$\tau_{33}$     & $c_1^{[0]}$     & $c_2^{[0]}$   & & &\\
\hline
  $1\times\frac{2\pi}{10}$   &  $2\times\frac{2\pi}{10}$  &  $3\times\frac{2\pi}{10}$  &  $0.46\times 2\pi$    &  $1.02\times 2\pi$  &  $1.53\times 2\pi$       &  $-1$       &  $1$ &&&\\
\hline
  $\omega_1$       &$\omega_2$    &  $\omega_3$        &  $l_1$         &  $l_2$      &  $l_3$     & $\tau_{12}$     & $\tau_{13}$    & $\tau_{23}$           &  $c_1$              &  $c_2$   \\
\hline
    $0.1981$&$2.1281$& $-2.8105$&$0.0942$&$0.0220$& $-0.1015$&$1.5832$&$-1.1454$&$1.2599$&$3.2383$&  $0.0428$\\
  \hline
    \hline
  $k_1$   &$k_2$    & $k_3$   &$\tau_{11}$     & $\tau_{22}$      &$\tau_{33}$     & $c_1^{[0]}$     & $c_2^{[0]}$   & & &\\
\hline
  $1\times\frac{2\pi}{10}$   &  $2\times\frac{2\pi}{10}$  &  $3\times\frac{2\pi}{10}$  &  $0.53\times 2\pi$    &  $0.75\times 2\pi$  &  $1.13\times 2\pi$       &  $1$       &  $1$ &&&\\
\hline
  $\omega_1$       &$\omega_2$    &  $\omega_3$        &  $l_1$         &  $l_2$      &  $l_3$     & $\tau_{12}$     & $\tau_{13}$    & $\tau_{23}$           &  $c_1$              &  $c_2$   \\
\hline
   $0.8089$&   $-1.0065$& $-0.1976$& $0$& $0$& $0$& $-1.5390$&$1.7911$&   $3.1734$&$36.8174$&$0$\\
  \hline
     \hline
  $k_1$   &$k_2$    & $k_3$   &$\tau_{11}$     & $\tau_{22}$      &$\tau_{33}$     & $c_1^{[0]}$     & $c_2^{[0]}$   & & &\\
\hline
  $1\times\frac{2\pi}{10}$   &  $2\times\frac{2\pi}{10}$  &  $3\times\frac{2\pi}{10}$  &  $0.53\times 2\pi$    &  $0.75\times 2\pi$  &  $1.13\times 2\pi$       &  $-1$       &  $1$ &&&\\
\hline
  $\omega_1$       &$\omega_2$    &  $\omega_3$        &  $l_1$         &  $l_2$      &  $l_3$     & $\tau_{12}$     & $\tau_{13}$    & $\tau_{23}$           &  $c_1$              &  $c_2$   \\
\hline
    $0.5147$&  $1.6120$&$-2.7811$& $-0.7670$& $-0.1002$&$0.8148$&$1.7865$&$-1.6037$&$1.2270$&$23.1536$&$-0.1155$\\
  \hline
\end{tabular}
\caption{$3$-periodic waves: examples with $v_0=0$}
\label{table31}
\end{table*}
\begin{table*}[!ht]\tiny
\center
\begin{tabular}{|c|c|c|c|c|c|c|c|c|c|c|}
\hline
  $k_1$   &$k_2$    & $k_3$   &$\tau_{11}$     & $\tau_{22}$      &$\tau_{33}$     & $c_1^{[0]}$     & $c_2^{[0]}$   & & &\\
\hline
  $1\times\frac{2\pi}{10}$   &  $2\times\frac{2\pi}{10}$  &  $3\times\frac{2\pi}{10}$  &  $0.67\times 2\pi$    &  $0.86\times 2\pi$  &  $1.02\times 2\pi$       &  $1$       &  $1$ &&&\\
\hline
  $\omega_1$       &$\omega_2$    &  $\omega_3$        &  $l_1$         &  $l_2$      &  $l_3$     & $\tau_{12}$     & $\tau_{13}$    & $\tau_{23}$           &  $c_1$              &  $c_2$   \\
\hline
 $1.4388$&$3.1394$& $7.9404$& $1.6866$&$2.0555$& $1.5475$& $2.1251$&    $1.2939$&    $2.7630$&   $24.0624$&     $0.9121$  \\
  \hline
  \hline
  $k_1$   &$k_2$    & $k_3$   &$\tau_{11}$     & $\tau_{22}$      &$\tau_{33}$     & $c_1^{[0]}$     & $c_2^{[0]}$   & & &\\
\hline
  $1\times\frac{2\pi}{10}$   &  $2\times\frac{2\pi}{10}$  &  $3\times\frac{2\pi}{10}$  &  $0.46\times 2\pi$    &  $1.02\times 2\pi$  &  $1.53\times 2\pi$       &  $-1$       &  $1$ &&&\\
\hline
  $\omega_1$       &$\omega_2$    &  $\omega_3$        &  $l_1$         &  $l_2$      &  $l_3$     & $\tau_{12}$     & $\tau_{13}$    & $\tau_{23}$           &  $c_1$              &  $c_2$   \\
\hline
 $1.3915$&  $3.3788$& $8.3344$& $1.9342$&$1.9653$&$1.5315$&$1.9341$&$1.2573$&   $2.8808$&  $6.0244$&  $0.9791$\\
  \hline
    \hline
  $k_1$   &$k_2$    & $k_3$   &$\tau_{11}$     & $\tau_{22}$      &$\tau_{33}$     & $c_1^{[0]}$     & $c_2^{[0]}$   & & &\\
\hline
  $1\times\frac{2\pi}{10}$   &  $2\times\frac{2\pi}{10}$  &  $3\times\frac{2\pi}{10}$  &  $0.53\times 2\pi$    &  $0.75\times 2\pi$  &  $1.13\times 2\pi$       &  $-1$       &  $1$ &&&\\
\hline
  $\omega_1$       &$\omega_2$    &  $\omega_3$        &  $l_1$         &  $l_2$      &  $l_3$     & $\tau_{12}$     & $\tau_{13}$    & $\tau_{23}$           &  $c_1$              &  $c_2$   \\
\hline
 $1.3608$&$2.4234$&$-3.9766$&$1.7988$&$2.3352$&$1.0781$&$2.0257$&$-1.9751$&$0.6098$&$44.3977$&$2.6950$\\
  \hline
\end{tabular}
\caption{$3$-periodic waves: examples with $v_0=1$}
\label{table32}
\end{table*}

\section{Conclusion and discussion}
\label{section5}
A numerical process of calculating $N$-periodic waves is presented and some numerical experiments are carried out with the coupled Ramani equation. The numerical results show that the process is efficient in calculating $1$-, $2$- and $3$-periodic waves. Here we give two remarks.

Firstly,  the numerical results are not unique since this is a nonlinear and over-determined system. For instance,  in the third and forth examples in Table~\ref{table31}, with the same given parameters but the different initial guess, we obtain two different $3$-periodic waves.

Secondly, there are many other soliton equations that can be transformed into the coupled bilinear KdV-type systems. For instance,  the Hirota-Satsuma coupled KdV equation\cite{hs}
\begin{eqnarray}
&&u_t=\frac{1}{4}u_{xxx}+3uu_x+3(-\phi^2+\omega)_x,\\
&&\phi_t=-\frac{1}{2}\phi_{xxx}-3u\phi_x,\\
&&\omega_t=\frac{1}{2}\omega_{xxx}-3u\omega_x.
\end{eqnarray}
can be transformed into the bilinear form
\begin{eqnarray}
&&(D_xD_t-\frac{1}{4}D_x^4-\frac{3}{4}D_z^2)f\cdot f=0,\\
&&D_z(D_t+\frac{1}{2}D_x^3)f\cdot f=0,
\end{eqnarray}
by the dependent variable transformation
\begin{equation}
u=(\mathrm{ln} f)_{xx},\quad \phi=\frac{1}{2}\frac{f_z}{f},\quad \omega=\frac{1}{2}\frac{f_{zz}}{f}
\end{equation}
where $z$ is an auxiliary variable,
and the  Camassa-Holm equation\cite{ch1,ch2}
\begin{equation}
u_t+2\kappa^2 u_x+3uu_x-u_{xxt}=2u_xu_xx+uu_{xxx}, \label{ch1}
\end{equation}
can be transformed into
\begin{eqnarray}
&&[D_y(D_t+2\kappa^3 D_y-\kappa^2 D_y^2D_t)\nonumber\\
&&\qquad\qquad\qquad+\frac{1}{3}\kappa^2D_t(D_{\tau}+D_y^3)]f\cdot f=0,\label{ch2}\\
&&D_y(D_{\tau}+D_y^3)f\cdot f=0.
\end{eqnarray}
with a so-called reciprocal transformation. Here $\tau$ is an auxiliary variable and coordinate $(y,t)$ is generated from the reciprocal transformation. See details in Ref.~\cite{ch2}.

Some discrete systems can also be transformed into this kind of bilinear equations. For example, the semi-discrete KdV equation given by Hirota and Ohta \cite{hirota3},
\begin{eqnarray}
&&\frac{4}{1+a^2u_n}\frac{d}{dt}u_n=\Delta^3Mu_n+6u_n\Delta Mu_n\nonumber\\
&&\qquad\qquad+a^2[\Delta M(u_n\Delta^2u_n)+3(\Delta Mu_n)(\Delta^2u_n)],\label{skdv1}
\end{eqnarray}
can be transformed into
\begin{eqnarray}
&&[2aD_z\sinh(\frac{D_n}{2})-\cosh(\frac{3D_n}{2})+\cosh(\frac{D_n}{2})]f\cdot f=0,\nonumber\\
&&\label{skdv2}\\
&&(4a^2D_t+3D_z)\sinh(\frac{D_n}{2})f\cdot f=D_z\sinh(\frac{3D_n}{2})f\cdot f,\label{skdv3}
\end{eqnarray}
with transformation
\begin{equation}
u_n=\frac{1}{a^2}(\frac{f_{n+2}f_{n-1}}{f_{n+1}f_n}-1).
\end{equation}

Our numerical process may also be applied to these soliton equations to study their $N$-periodic wave solutions if some additional terms with arbitrary integral constants are introduced in these bilinear forms.

\section*{Acknowledgements}
This work was partially supported by the National Natural Science Foundation of China (Grant no. 11401546, 11601237, 11571358, 11331008), the Natural Science Foundation of Jiangsu Province Colleges and Universities (Grant no. 16KJB110014) and Jiangsu Planned Projects for Postdoctoral Research Funds(No.1601054A).

\end{document}